\documentclass[12pt,letterpaper]{article}
\usepackage{booktabs} 

\usepackage{subcaption} 
\usepackage{amsmath}
\usepackage{mathtools}
\usepackage[longnamesfirst]{natbib}
\usepackage{bm} 
\usepackage{amsthm,amssymb}

\theoremstyle{plain} 

\newtheorem{proposition}{Proposition}
\newtheorem{definition}{Definition}

\newtheorem{corollary}{Corollary}
\theoremstyle{definition} 

\usepackage[labelfont=bf]{caption}
\usepackage{pdflscape}

\usepackage{graphicx,subcaption}

\usepackage{xpatch} 
\xpatchcmd{\proof}{\itshape}{\normalfont\proofnamefont}{}{} 
\newcommand{\proofnamefont}{\bfseries} 

\usepackage[title]{appendix} 
\usepackage{threeparttable, booktabs, siunitx} 
\usepackage{dcolumn}
\usepackage{enumitem}

\usepackage{lipsum}
\usepackage{setspace}
\usepackage[dvipsnames]{xcolor}

\usepackage{hyperref}
\definecolor{internationalkleinblue}{rgb}{0.0, 0.18, 0.65}
\definecolor{blue(pigment)}{rgb}{0.2, 0.2, 0.6}
\hypersetup{colorlinks=true, citecolor=internationalkleinblue, linkcolor=internationalkleinblue, urlcolor  = black}
\usepackage[letterpaper, margin=1.2in]{geometry} 
\usepackage{longtable}

\usepackage{multirow} 
\usepackage{tocloft}
\usepackage{dirtytalk} 
\usepackage{dcolumn}
\usepackage{tikz}

\makeatletter
\newcommand*{\centerfloat}{%
  \parindent \z@
  \leftskip \z@ \@plus 1fil \@minus \textwidth
  \rightskip\leftskip
  \parfillskip \z@skip}
\makeatother

\title{\vspace{-1.0cm} \textsf{On the Instability of Fractional Reserve Banking}\footnote{This paper is a revised version of Chapter 2 of my PhD dissertation at the University of Missouri. I am deeply indebted to my advisor, Chao Gu, for her guidance at all stages of this research project. I am grateful to my committee members, Joseph Haslag, Aaron Hedlund, and Xuemin (Sterling) Yan for helpful comments. I also thank Costas Azariadis, Guido Menzio, Christian Wipf and seminar participants at the 2020 MVEA Conference, 2021 EGSC, and 2021 MEA Annual Meeting for useful feedback and discussion. The views expressed in this paper are solely those of the author. All remaining errors are mine.}}

\author{\textsc{Heon Lee}\footnote{Contact: \href{mailto:heonlee68@gmail.com}{heonlee68@gmail.com}} \\ University of Missouri}

\begin{document}
\maketitle
\onehalfspacing

\begin{abstract}
\noindent 
This paper develops a dynamic monetary model to study the (in)stability of the fractional reserve banking system. The model shows that the fractional reserve banking system can endanger stability in that equilibrium is more prone to exhibit endogenous cyclic, chaotic, and stochastic dynamics under lower reserve requirements, although it can increase consumption in the steady-state. Introducing endogenous unsecured credit to the baseline model does not change the main results. The calibrated exercise suggests that this channel could be another source of economic fluctuations. This paper also provides empirical evidence that is consistent with the prediction of the model.
\end{abstract}

\newpage
\onehalfspacing
\begingroup
\addtolength\leftmargini{0.04in}
\begin{quote}
Motivated partly by a desire to avoid such [excessive] price-level fluctuations and possible Wicksellian price-level indeterminacy, quantity theorists have advocated legal restrictions on private intermediation. ... Thus, for example, Friedman (1959, p. 21) ... has advocated 100 percent reserves against bank liabilities called demand deposit. 
\hspace*{\fill}\textbf{\cite{sargent1982real}}
\end{quote}
\endgroup
\onehalfspacing

\section{Introduction}

There have been claims that fractional reserve banking is an important cause of boom-bust cycles, based on the notion that  banks create excess credit under fractional reserve banking. (e.g., \citealp*{fisher1935100}; \citealp*{von1953theory};  \citealp*{minsky1957monetary}; \citealp*{minsky1970financial}). For instance, \cite{fisher1935100} views fractional reserve banking as one of several important factors in explaining economic fluctuations. Others believe that this is a primary cause of boom-bust cycles. According to \cite{von1953theory}, the overexpansion of bank credit as a result of fractional reserve banking is the root cause of business cycles. \cite{minsky1970financial} claims that economic booms and structural characteristics of the financial system, such as fractional reserve banking, can result in an economic collapse even when fundamentals remain unchanged.

This idea leads to policy debates on fractional reserve banking. Earlier examples include Peel's Banking Act of 1844 and the Chicago plan of banking reform with a 100\% reserve requirement proposed by Irving Fisher, Paul Douglas, and others in 1939. Later, \cite{friedman1959program} supported this banking reform, whereas \cite{becker1956free} took the opposite position of supporting free banking with 0\% reserve requirement.\footnote{\cite{sargent2011draw}  provides a novel review of the historical debates between narrow banking and free banking as tensions between stability versus efﬁciency.} Recently in 2018, Switzerland had a referendum of 100\% reserve banking, which was rejected by 75.72\% of the voters. The referendum aimed at making money safe from crisis by constructing full-reserve banking.\footnote{The official title of the referendum was \textit{the Swiss federal popular initiative ``for crisis-safe money: money creation by the National Bank only! (Sovereign Money Initiative)"} and also titled as ``\textit{debt-free money}."} Whereas the debate on whether a fractional reserve banking system is inherently unstable has been an important policy discussion since a long time ago, the debate has never stopped.

This paper examines the instability of fractional banking by answering the following questions:  (i) Can fractional reserve banking be inherently volatile  even if we shut down the stochastic component of the economy? (ii) If so, under what condition can fractional reserve banking generate endogenous cycles without the presence of exogenous shocks and changes in fundamentals? To assess the claim that fractional reserve banking causes business cycles, this paper constructs a model of money and banking that captures the role of fractional reserve banking. 

In the model, each agent faces an idiosyncratic liquidity shock. Banks accept and issue deposits and extend loans to provide risk-sharing among the depositors. The bank makes loans by creating deposit money, and its lending deposit money creation is constrained by the reserve requirement. At equilibrium, the real balance of money is determined by two factors: storage value and liquidity premium. The storage value increases with the future value of money. However, the liquidity premium, which is the marginal value of money's liquidity function, decreases as money becomes more abundant. When the liquidity premium dominates the storage value, the economy can exhibit endogenous fluctuations. Fractional reserve banking amplifies the liquidity premium because it allows the bank to create inside money through lending. Due to this amplified liquidity premium, the fractional reserve banking system is more prone to endogenous cycles.

In the baseline model, lowering the reserve requirement increases consumption in the steady state. However, lowering the reserve requirements can induce two-period cycles as well as three-period cycles, which implies the existence of periodic cycles of all order and chaotic dynamics. This also implies it can induce sunspot cycles. This result holds in the extended model with unsecured credit. The model also can deliver a self-fulfilling bubble burst. It is worth noting that the full reserve requirement does not  necessarily exclude the possibility of endogenous cycles. However, the economy will be more susceptible to cycles with lower reserve requirement.\footnote{\cite{gu2019instability} show that introducing banks to the economy could induce instability in various settings which is in line with this result.}

This paper departs from previous works in two ways. First, in contrast to the previous works on banking instability, which mostly focus on bank runs following the seminal model by \cite{diamond1983bank}, this paper focuses on the volatility of real balances of  money. It is another important focal point of banking instability because recurring boom-bust cycles associated with banking are probably be more prevalent than bank runs. Second, the approach here differs from a traditional approach to economic fluctuations with financial frictions. To understand economic fluctuations, there are two major points of view. The first one is that economic fluctuations are driven by exogenous shocks disturbing the dynamic system, and the effects of exogenous shocks shrink over time as the system goes back to its balanced path or steady-state. The second one is that they instead reflect an endogenous mechanism that produces boom-bust cycles. While there has been a lot of work on the role of financial friction in the business cycles including \cite{kiyotaki1997credit}, \cite{bernanke1999financial}, and \cite{gertler2011model}, most of them focused on the first approach, in which all economic fluctuations are caused by exogenous shocks and the financial sectors only serve as an amplifier. This paper, however, takes the second approach and focuses on whether the endogenous cycles arise in the absence of the stochastic components of the economy.

To evaluate the main prediction from the theory that fractional reserve banking induces excess volatility, I test the relationship between the required reserves ratio and the volatilty in real balance using cointegrating regression. A significant negative relationship between the two variables are found, and the results are robust to different measures of inflation and different frequency of time series. Both theoretical and empirical evidence indicate a link between the reserve requirement and the (in)stability.  \\

\noindent\textbf{Related Literature } This paper builds on \cite{berentsen2007money}, who introduce financial intermediaries with enforcement technology to \cite{lagos2005unified} framework. The approach to introduce unsecured credit to the monetary economy is related to \cite{lotz2016money} and \cite{gu2016money} which are based on the earlier work by \cite{kehoe1993debt}.  

This paper is related to the large literature on fractional reserve banking. \cite{freeman1991inside} and \cite{freeman2000monetary} develop general equilibrium models that explicitly capture the role of fractional reserve banking. Using those models, they explain the observed relationships between key macroeconomic variables over business cycles. \cite{chari2014social} study an economy where private agents have incentives to establish fractional reserve banking as an alternative payment system. This alternative system is inherently fragile because it is susceptible to socially costly bank runs. They study the conditions under which the social benefits of fractional reserve banking can exceed its social costs which crucially depend on communication technologies. For recent work, \cite{wipf2020should} studies the welfare implications of fractional reserve banking in a New Monetarist economy with imperfect competition and identifies the conditions under which fractional reserve banking can be welfare-improving compared to narrow banking. \cite{andolfatto2020money} integrate \cite{diamond1997liquidity} into \cite{lagos2005unified} to provide a model in which fractional reserve banking emerges endogenously and a central bank can prevent bank panic as a lender of last resort. Whereas many previous work on instability focuses on bank runs or societal value at the steady state, this paper studies a different type of instability in the sense that fractional reserve banking induces endogenous monetary cycles.  

This paper is also related to the large literature on endogenous fluctuations, chaotic dynamics, and indeterminacy that have been surveyed by \cite{brock1988nonlinearity}, \cite{baumol1989chaos}, \cite{boldrin1990equilibrium}, \cite{scheinkman1994self}  and \cite{benhabib1999indeterminacy}. For a model of bilateral trade, \cite{gu2013endogenous} show that credit markets can be susceptible to endogenous fluctuations due to limited commitment.  Using a continuous-time New Monetarist economy, \cite{rocheteau2023endogenous} show that asset liquidity can be a source of price volatility when assets have a non-positive intrinsic value. \cite{altermatt2023general} study economies with multiple liquid assets and show that liquidity considerations could imply endogenous fluctuations as self-fulfilling prophecies. \cite{gu2019instability} show that introducing financial intermediaries to an economy can engender instability in the sense that endogenous cycles are more likely to emerge in the presence of financial intermediaries. They demonstrate this in four distinct setups that capture various functions of banking. The model in this paper is closely related to that of \cite{gu2019instability} in the sense that both papers study the role of banking in instability. However, this paper goes further by focusing on the fractional reserve banking system. One focal point is that, rather than treating banks as mere intermediaries, in this paper, the bank creates deposit money (inside money) by making loans. Furthermore, this paper establishes exact thresholds of reserve requirements under which the equilibrium can exhibit endogenous cycles and chaotic dynamics. \\

The rest of the paper is organized as follows. Section \ref{sec:model} constructs the baseline search-theoretic monetary model. Section \ref{sec:results} provides main results. Section \ref{sec:credit} introduces unsecured credit. Section \ref{sec:quant} calibrates the model to quantify the theory. Section \ref{sec:empiric} discusses the empirical evaluation of the model's prediction.  Section \ref{sec:conclusion} concludes. 

\section{Model}
\label{sec:model}

The model is based on \cite{lagos2005unified} with banking as in \cite{berentsen2007money}. Time is discrete and infinite. In each period, three markets convene sequentially. First, a centralized financial market (FM), followed by a decentralized goods market (DM), and finally a centralized goods market (CM). The FM and CM are frictionless. The DM is subject to search frictions, anonymity, and limited commitment. Therefore, a medium of exchange is needed to execute trades.

There is a continuum of agents who produce and consume perishable goods. At the beginning of the FM, a preference shock is realized: With probability $\sigma$, an agent will be a buyer in the following DM and with probability $1-\sigma$, she will be a seller. The buyers and the sellers randomly meet and trade bilaterally in the DM. Agents discount their utility each period by $\beta$. Within-period utility is represented by 
$$\mathcal{U}= U(X)-H+u(q)-c(q),$$
where $X$ is the CM consumption, $H$ is the CM disutility from production, and $q$ is the DM consumption. As standard $U'$, $u'$, $c'>0$, $U''$, $u''<0$, $c''\geq0$, and $u(0) = c(0)=0$.
The CM consumption good $X$ is produced one-for-one with $H$, implying the real wage is 1. The efficient consumption in CM and DM is $X^*$ and $q^*$ that solve $U'(X^*)=1$ and $u'(q^*)=c'(q^*)$, respectively. 

\begin{figure}[t!]\label{fig:balance_sheet}
\centering
\includegraphics[width=15cm,height=6.5cm]{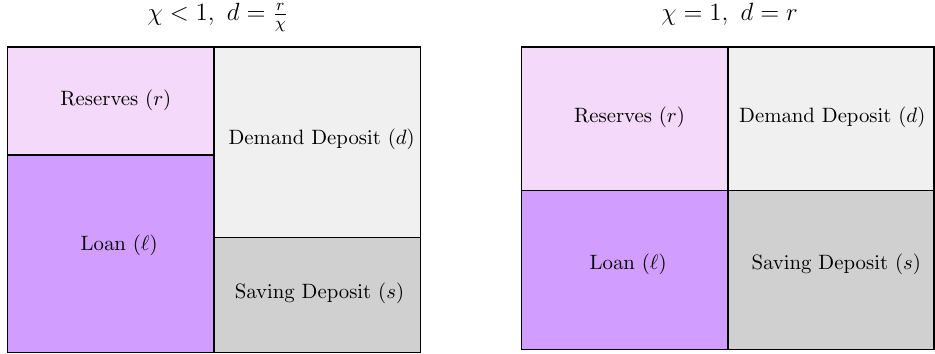}
\caption{Bank's Balance Sheet: fractional reserve banking vs. full reserve banking}
\end{figure}

There is a representative bank that accepts and issues deposits and lends loans in the FM. In the FM, an agent can borrow money from the bank with a promise to repay the money in the subsequent CM at a nominal lending rate $i_l$. There are two kinds of deposits: demand deposits and saving deposits. The agent can deposit her fiat money into the bank's savings deposit and receive money in the subsequent CM at a nominal deposit rate. The agent can also hold her balance in a demand deposit. The demand deposit does not pay interest, while the interest rate on saving deposits is $i_s$. In addition to that, the demand deposit can be used as a means of payment in DM trade, whereas the savings deposit cannot. When the bank lends loans, it creates demand deposits, and the bank's issuance of demand deposits is subject to a reserve requirement, $\chi$. Figure 1 illustrates the bank's balance sheet identity given the reserve requirement. The banking market is perfectly competitive, and the bank can enforce the repayment of loans at no cost. Lastly, there is a central bank that controls the fiat money supply $M_t$. Let $\gamma$ be the growth rate of the fiat money stock. Changes in the fiat money supply are accomplished by lump-sum transfers if $\gamma>0$ and by lump-sum taxes if $\gamma<0$.

\subsection{Agent's Problem}
Let $W_t$, $G_t$, and $V_t$ denote the agent's value function in the CM, FM, and DM, respectively, in period $t$. There are two payment instruments for the DM transaction: fiat money and demand deposit. However, buyers and sellers do not discriminate between these instruments in the DM transaction because agents treat them as the same 'money.' The agent's state variables in the CM are $a_t$, $s_t$, and $\ell_t$, where $a_t=m_t+d_t$, $s_t$ is a saving deposit, $\ell_t$ is a loan borrowed from the bank, $m_t$ is fiat money (outside money) issued by the central bank, and $d_t$ is a demand deposit (inside money) issued by the bank, The state variable $a_t$ represents the agent's nominal balance of liquid assets, which can be used for transactions in the DM. I will allow the agents to use unsecured credit as a means of payment in the next section. An agent entering the CM with nominal balance $a_t$, saving deposit $s_t$, and loan $\ell_t$ solves the following problem:
\begin{equation}
\label{eq:CM}
\begin{split}
W_{t}(a_{t},s_{t},\ell_{t})=&\max_{X_{t},H_{t},\hat{m}_{t+1}} U(X_t)-H_t+\beta G_{t+1}(\hat{m}_{t+1}) \\
\text{s.t. } \phi_{t}\hat{m}_{t+1}+X_{t}= & H_{t}+T_{t}+\phi_{t} a_{t}+(1+i_{s,t})\phi_{t} s_{t}-(1+i_{l,t})\phi_{t}\ell_{t},
\end{split}
\end{equation}
where $T_{t}$ is the lump-sum transfer (or tax if it is negative), $i_{s,t}$ is the savings deposit interest rate, $i_{l,t}$ is the loan interest rate, $\phi_t$ is the price of money in terms of the CM goods, and $\hat{m}_{t+1}$ is the money balance carried to the FM where banks take deposits and makes loans. The first-order conditions (FOCs) result in $X_{t}=X^*$ and 
\begin{align}
\label{eq:MVM1}
\phi_{t}=\beta G'_{t+1}(\hat{m}_{t+1}),
\end{align}
where $G'_{t+1}(\hat{m}_{t+1})$ is the marginal value of an additional unit of money taken into the FM of period $t+1$. The envelope conditions are 
$$ \frac{\partial W_{t}}{\partial a_{t}}=\phi_{t}, \qquad \frac{\partial W_{t}}{\partial s_{t}}=\phi_{t}(1+i_{s,t}),  \qquad \frac{\partial W_{t}}{\partial \ell_{t}}=-\phi_{t}(1+i_{l,t}),$$
implying $W_t$ is linear in $m_t$, $s_t$, and $\ell_t$.

The value function of an agent at the beginning of FM is 
\begin{equation}
G_{t}(m)=\sigma G_{b,t}(m)+ (1-\sigma)G_{s,t}(m),
\end{equation} 
where $G_{j\in\{b,s\},t}$ is the value function of type $j$ agent in the FM. Agents choose their deposit balances $d_j$, $s_j$ and loan $\ell_j$ based on the realization of their types in the following DM and they can acquire demand deposits by borrowing loans from the bank. 

The value function $G_{j,t}$ can be written as 
\begin{equation}
\label{eq:FM}
\begin{split}
G_{j,t}(m)=&\max_{d_{j,t},s_{j,t},\ell_{j,t}} V_{j,t}(m+d_{j,t}-s_{j,t},s_{j,t},\ell_{j,t}) \\
\text{s.t. } \quad  &s_{j,t}\leq m, \text{ and }  d_{j,t}=  \ell_{j,t}
\end{split}
\end{equation}
where $V_{j,t}$ is the value function of type $j$ agent in the DM. The FOCs are
\begin{align}
\frac{\partial V_{j,t}}{\partial \ell_{j,t}}&\leq0 \label{eq:FM_foc1}\\ 
\frac{\partial V_{j,t}}{\partial s_{j,t}}&-\lambda_s\leq0
\label{eq:FM_foc2} 
\end{align}
where $\lambda_s$ is the Lagrange multiplier for $s_{j,t}\leq m$.

The terms of trade in the DM are determined by an abstract mechanism that is studied in  \cite{gu2016monetary}. The buyer must pay $p=v(q)$ to the seller to get $q$ where $v(q)$ is some payment function satisfying $v'(q)>0$ and $v(0)=0$. As shown in \cite{gu2016monetary}, if the trading protocol satisfies four common axioms, then the terms of trade can be written in the following form.
\begin{align}\label{eq:term_trade}
 p  = 
  \begin{dcases} 
   z    & \text{if } z < p^*\\ 
   p^*    & \text{if } z \geq p^*
  \end{dcases}
\qquad  q  = 
  \begin{dcases} 
   v^{-1}(z)    & \text{if } z < p^*\\ 
   q^*    & \text{if } z \geq p^*,
  \end{dcases} 
\end{align} 
where $p^*$ is the payment required to get efficient consumption $q^*$, and $z$ is the total liquidity, $(m-s+d)\phi$, held by the buyer. Many standard mechanisms, such as Kalai and generalized Nash bargaining, are consistent with this specification.

With probability $\alpha$, a buyer meets a seller in the DM while a seller meets a buyer with probability $\alpha_s$. Since the CM value function is linear, the DM value function for the buyer can be written as
\begin{equation}\label{eq:dm_buyer}
V_{b,t}(m_t+d_{b,t}-s_{b,t},s_{b,t},\ell_{b,t})=\alpha [u(q_t)-p_t]+W(m_t+d_{b,t}-s_{b,t},s_{b,t},\ell_{b,t}),
\end{equation}
where $p_t\leq (m_t-s_{b,t}+d_{b,t})\phi_t$. Assuming 
interior solution,  differentiating $V_{b,t}$ yields 
$$\frac{\partial V_{b,t}}{\partial m}=\frac{\partial V_{b,t}}{\partial d}=\phi_t [\alpha \lambda(q_t)+1], \qquad \frac{\partial V_{b,t}}{\partial s}=\phi_t [-\alpha \lambda(q_t)+i_{s,t}], \qquad \frac{\partial V_{b,t}}{\partial \ell}=\phi_t [\alpha \lambda(q_t)-i_{l,t}],$$
where $\lambda(q)=u'(q)/v'(q)-1$ if $p^*>z$ and $\lambda(q)=0$ if $z\geq p^*$. Combining the buyer's FOCs in the FM and the derivatives of $V_b$ yields
\begin{align}
 \phi i_{s,t} -\phi\alpha\lambda(q_t) -\lambda_s &\leq 0, ``="0 \text{ iff } s_{b,t}>0            \\
 -\phi i_{l,t} +\phi\alpha\lambda(q)  &\leq 0     , ``="0 \text{ iff } \ell_{b,t}>0.\label{eq:loan_rate}   
\end{align} 
A seller's DM value function is
\begin{equation}\label{eq:dm_seller}
V_{s,t}(m_t+d_{s,t}-s_{s,t},s_{s,t},\ell_{s,t})=\alpha_s [p_t-c(q_t)]+W_{t}(m_t+d_{s,t}-s_{s,t},s_{s,t},\ell_{s,t}).
\end{equation}
where $d_{s,t}=\ell_{s,t}$. Differentiating $V_{s,t}$ after substituting the constraint yields
$$\frac{\partial V_{s,t}}{\partial m_{t}}=\frac{\partial V_{s,t}}{\partial d_{t}}=\phi_{t},  \qquad \frac{\partial V_{s,t}}{\partial s_{t}}=\phi_{t}i_{s,t},  \qquad \frac{\partial V_{s,t}}{\partial \ell_{t}}=-\phi_{t}i_{l,t}.$$
Similar to the buyer's case, combining the seller's FOCs in the FM and the first-order derivatives of $V_{s,t}$ yields
\begin{align}
 \phi_{t} i_{s,t} -\lambda_s &\leq 0, ``="0 \text{ iff } s_{s,t}>0   \\
-\phi_{t} i_{l,t}  &\leq 0, ``="0 \text{ iff } \ell_{s,t}>0.
\end{align}

One can show that buyers do not deposit into savings deposits and sellers always deposit into savings deposits, whereas buyers always borrow loans, but sellers do not. This is because the buyer needs liquidity to trade for $q$ in the DM but the seller does not. Formally, for $m>0$, we have $\partial V_{s,t}/\partial s_{b,t}<\partial V_{s,t}/\partial s_{s,t}=0$ and $\partial V_{s,t}/\partial \ell_{s,t}<\partial V_{b,t}/\partial \ell_{b,t}=0$  because 
\begin{align}
0&=\overbrace{i_{s,t} -\lambda_d/\phi_t}^{\partial V_{s,t}/\partial s_{s,t}} > \overbrace{i_{s,t}-\lambda_s /\phi_t -\alpha\lambda(q_t)}^{\partial V_{b,t}/\partial s_{b,t}}   \\
0&=\underbrace{-\phi_t i_{l,t} +\phi_t\alpha\lambda(q_t)}_{\partial V_{b,t}/\partial \ell_{b,t}} >\underbrace{-\phi_t i_{l,t}}_{\partial V_{s,t}/\partial \ell_{s,t}}
\end{align}
implying $i_{l,t}=\alpha\lambda(q_t)$, $s_{s,t}=m$, $s_{b,t}=0$, $\ell_{s,t}=0$, and $\ell_{b,t}>0$ as long as $\lambda(q_t)>0$.

Using the above results, we can rewrite the value functions in the FM as follows:
\begin{align}
G_{b,t}(m_t)&=\alpha[u(q_t)-p_t]+W(m_t+d_{b,t},0,\ell_{b,t}) \\
G_{s,t}(m_t)&=\alpha_s[p_t-c(q_t)]+W(0,m_{t},0)
\end{align}
where $q_t=v^{-1}(p_t)$, $d_{b,t}=\ell_{b,t}$ and $p_t=\min\{p^*, (m_t +d_{b,t})\phi_{t}\}$. Take derivative of $G_{j,t}(m_t)$ with respect to $m_t$ to get
\begin{align}
G_{b,t}'(m_t)&=\phi_{t}+\phi_{t} \alpha\lambda(q_{t})  \\
G_{s,t}'(m_t)&=\phi_{t}+\phi_{t} i_{s,t}.
\end{align}
Since $G'_{t}(m_{t})= \sigma G'_{b,t}(m_{t})+(1-\sigma)G'_{s,t}(m_{t})$, we have the following:
\begin{equation}\label{eq:MVM2}
G'_{t}(m_{t})=\phi_{t}\sigma [1+\alpha \lambda(q_{t})]+\phi_{t}(1-\sigma)\left(1+i_{s,t}\right).
\end{equation} 
Combine (\ref{eq:MVM1}) and (\ref{eq:MVM2}) to get the Euler equation
\begin{align}
\label{eq:euler}
\phi_{t}= 
\begin{dcases} 
\phi_{t+1}\beta\left[ \sigma\left\{1+\alpha\lambda(q_{t+1})\right\}+(1-\sigma) (1+i_{s,t+1})\right] &\text{ if } z_{t+1}< p^* \\
\phi_{t+1}\beta            &\text{ if } z_{t+1}\geq p^*,
\end{dcases}
\end{align}  
where $q_{t+1}=v^{-1}(z_{t+1})$ and $z_{t+1}=(m_{t+1}+d_{b,t+1})\phi_{t+1}$ 

\subsection{Bank's Problem and Equilibrium}

A representative bank accepts saving deposits $s_t$, issues demand deposits $d_t$, and makes loans $\ell_t$. The bank is required to hold reserves equal to $\chi d_t$. The bank pays a nominal interest rate of $i_{s,t}$ to depositors for their savings deposits, while borrowers must repay their loans with a nominal interest rate of $i_{l,t}$. Demand deposits do not pay interest. The central bank sets the reserve requirement $\chi$. Given these conditions, the representative bank solves the following profit maximization problem.
\begin{equation}
\label{eq:bank}
\begin{split}
\max_{r_t,d_t,\ell_t,s_t} \quad &(1+i_{l,t})\ell_{t}+ r_t-d_{t}-(1+i_{s,t})s_t \\
\text{ s.t. } \quad \ell_{t}+r_t &= d_{t} + s_t , \quad d_t=\ell_t \quad \text{ and } \quad r_t\geq \chi d_{t}
\end{split}
\end{equation}
In the first constraint, the balance sheet identity, the left-hand side represents assets, which include reserves and loans, while the right-hand side represents liabilities, which include demand deposits and saving deposits. The second constraint simply states that banks create demand deposits by making loans, and the last constraint is the reserve requirement constraint. By substituting the two constraints, the bank's problem can be written as: \begin{equation}
\max_{\ell_t,r_t} \quad (1+i_{l,t})\ell_{t} + r_t-\ell_{t}-(1+i_{s,t})r_t \quad \text{ s.t. } r_t\geq \chi \ell_{t}
\end{equation}
The FOCs for the bank's problem are
\begin{align}
&0=i_{l,t}-\lambda_\chi\chi   \\
&0=-i_{s,t}+\lambda\chi
\end{align}
With the binding reserve requirement constraint, we have 
\begin{equation}
\label{eq:bank_eq}
i_{l,t}= \chi i_{s,t}.
\end{equation}

Given the bank's problem and the agent's problem, we can define an equilibrium as follows: \spacing{1}
\begin{definition}
\label{def:eqm}
Given $(\gamma,\chi)$, an equilibrium consists of sequences of prices $\{\phi_t,i_{l,t},i_{s,t}\}_{t=0}^{\infty}$, real balances $\{m_t,\ell_{b,t},\ell_{s,t},d_{b,t},d_{s,t},s_{b,t},s_{s,t}\}_{t=0}^{\infty}$, and allocations $\{q_t,X_t\}_{t=0}^{\infty}$ satisfying the following:
\begin{itemize}
\item Agents solve CM, FM and DM problems: (\ref{eq:CM}) and (\ref{eq:FM})
\item The terms of trade in the DM satisfy (\ref{eq:term_trade}), (\ref{eq:dm_buyer}) and (\ref{eq:dm_seller})
\item A representative bank solves its profit maximization problem: (\ref{eq:bank})
\item Markets clear in every period:
\begin{enumerate}
\item Deposit Markets: $\sigma d_{b,t}+(1-\sigma)d_{s,t}=d_{t}$ \text{ and } $\sigma s_{b,t}+(1-\sigma)s_{s,t}=s_{t}$ 
\item Loan Market: $\sigma\ell_{b,t}+(1-\sigma)\ell_{s,t}=\ell_{t}$
\item Money Market: $m_t=M_t$
\end{enumerate}
\item Transversality condition: $\lim_{t\rightarrow\infty}\beta^t \phi_t m_t=0$
\end{itemize}
\end{definition}
\onehalfspacing

The next step is to characterize the equilibrium. Given the agents' and the bank's problem with the binding reserve requirement constraint, we have $\ell_{b,t}=d_{b,t}=(1-\sigma)m_t/(\sigma\chi)$. 
Combine equations (\ref{eq:loan_rate}), (\ref{eq:euler}), and (\ref{eq:bank_eq}), and use the equilibrium condition $m_{t+1}=M_{t+1}$ to get
\begin{align}
\label{eq:dif}
\phi_{t}= 
\begin{dcases} 
\phi_{t+1}\beta\left[\frac{1-\sigma+\sigma\chi}{\chi}\alpha\lambda\circ v^{-1} (z_{t+1})+1\right] &\text{ if } z_{t+1}< p^* \\
\phi_{t+1}\beta            &\text{ if } z_{t+1}\geq p^*,
\end{dcases}
\end{align}  
where $z_{t+1}=\phi_{t+1} M_{t+1}(1-\sigma+\sigma\chi)/\sigma\chi$. Then multiplying both sides of (\ref{eq:dif}) by $M_{t}(1-\sigma+\sigma\chi)/\sigma\chi$ allows us to reduce the equilibrium condition to one difference equation of real balances $z$:
\begin{align}
\label{eq:mdynamics}
z_{t}=f(z_{t+1}) \equiv 
\frac{z_{t+1}}{1+i}\left[\frac{1-\sigma+\sigma\chi}{\chi}\alpha L(z_{t+1})+1\right],
\end{align}  
where $(1+i)\equiv\gamma/\beta$ and $L(z)\equiv\lambda\circ v^{-1}(z)$ is the liquidity premium.\footnote{In the stationary equilibrium, $i=\gamma/\beta-1$ is the nominal interest rate.} When $z_{t+1}\geq p^*$, $q_{t+1}=q^*$ because the buyer has sufficient liquidity to buy $q^*$. In this case, the liquidity is abundant. When $z_{t+1}< p^*$, $q_{t+1}=v^{-1}(z_{t+1})<q^*$ because the buyer does not have enough liquidity to buy $q^*$. In this case, the liquidity is scarce.

\section{Results}
\label{sec:results}

\begin{figure}\centerfloat 
  \begin{subfigure}{0.55\textwidth}
    \includegraphics[width=\linewidth]{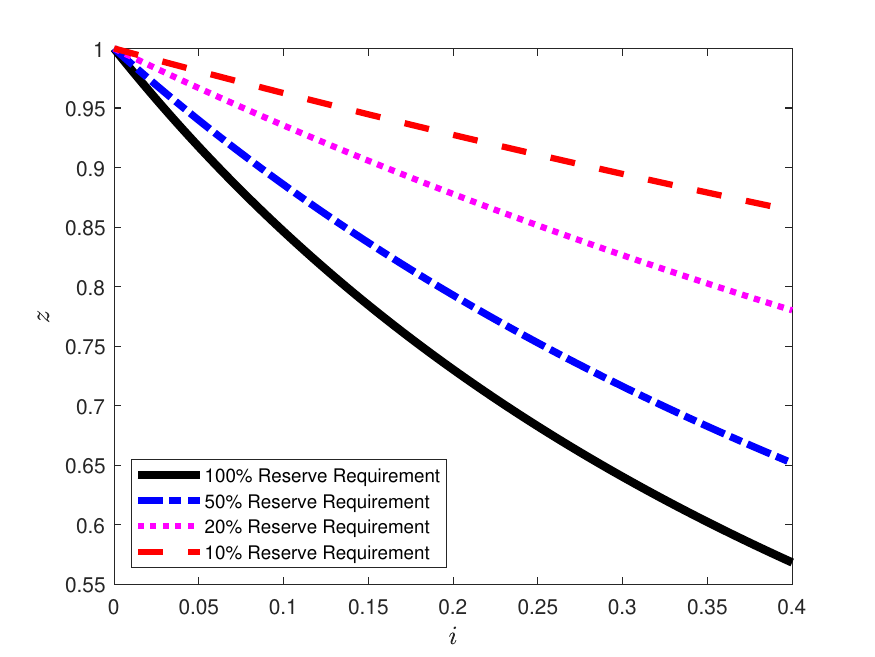}
    \caption{Stationary Monetary Equilibrium} \label{fig:1a}
  \end{subfigure}%
  \begin{subfigure}{0.55\textwidth}
    \includegraphics[width=\linewidth]{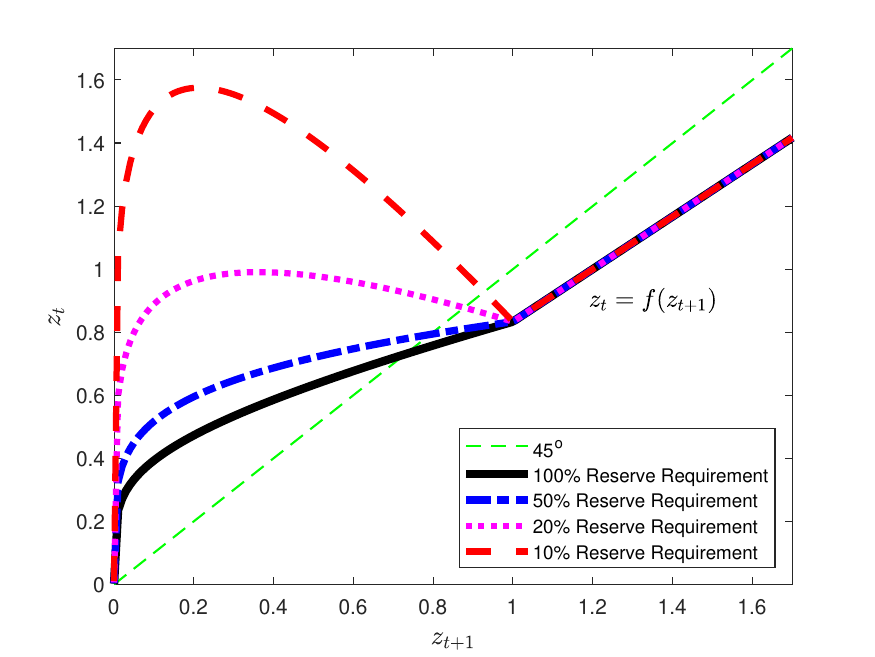}
    \caption{Dynamic Monetary Equilibrium} \label{fig:1b}
  \end{subfigure}
  \caption{Monetary Equilibrium} \label{fig:mss}
\end{figure}

This section establishes key results. Before starting a discussion on dynamics, consider stationary equilibria which are defined as fixed points that satisfy $z=f(z)$. There always exists a non-monetary equilibrium with $z=0$. A unique solution of monetary stationary equilibrium $z_s>0$ exists and solves
\begin{equation}\label{eq:mss}
\chi i=(1-\sigma+\sigma\chi)\alpha L(z_s)
\end{equation}
when $i<\bar\iota$ where $\bar\iota=\alpha(1-\sigma+\sigma\chi)L(0)/\chi$. Nash and Kalai bargaining provide simple examples for $\bar\iota$. Under the Inada condition $u'(0)=\infty$, with Kalai, $\bar\iota=\theta\alpha(1-\sigma+\sigma\chi)/\chi(1-\theta)$; whereas with Nash bargaining, $\bar\iota=\infty$. 

Since $L'(z)<0$ and $v'(q)>0$ (see \citealp*{gu2016monetary}), the following result holds: 
\begin{proposition}
\label{prop:welfare}
In the stationary equilibrium, lowering $i$ or lowering $\chi$ increases $q$.
\end{proposition}
\begin{proof}
See Appendix \ref{sec:proof}.
\end{proof}
Figure \ref{fig:1a} plots $z$ against $i$. It shows downward-sloping money demand in the stationary equilibrium given the reserve requirement. Lowering the reserve requirement increases $z$ because it allows a bank to create more liquidity in the economy which increases $q$ as well. 
 
The dynamics of monetary equilibrium are characterized by $f(z_{t+1})$ from equation (\ref{eq:mdynamics}).  The first term, $z_{t+1}/(1+i)$ on the right-hand side, reflects the store of value, which is monotonically increasing in $z_{t+1}$. The second term  $(1-\sigma+\sigma\chi)\alpha L(z_{t+1})/\chi+1$, reflecting the liquidity premium, is decreasing in $z_{t+1}$. Because $f'(z_{t+1})$ depends on both terms, $f(z_{t+1})$ is nonmonotone in general. Figure \ref{fig:1b} provides an example. In this example, as the reserve requirement decreases, the equation (\ref{eq:mdynamics}) is more likely to have the backward bending feature. Lowering the reserve requirement amplifies the liquidity premium, as it enables banks to generate more liquidity through lending. This amplification of liquidity enhances the backward-bending feature, potentially leading to endogenous cycles.

The standard treatment for showing the existence of an endogenous cycle is $f'(z_s)<-1$ (see \citealp*{azariadisintertemporal}). In this case, the economy can exhibit a two-period cycle with $z_1<z_s< z_2$ which can be either $z_2<p^*$ or $z_2\geq p^*$. However, without further assumptions, we cannot determine the conditions under which this can occur. For illustration, let's take the derivative of (\ref{eq:euler}) with respect to $z_{t+1}$ and evaluate it at $z_{t+1} = z_s$. We obtain the following expression:
\begin{equation}\label{eq:fprime1}
f'(z_{s})= 
\frac{1}{1+i}\left[\frac{\alpha(1-\sigma+\sigma\chi)}{\chi}\left\{L'(z_{s})z_{s}+L(z_{s}) \right\}+1\right]
\end{equation}
As $L'(z_{s})$ is not explicitly defined here, we are unable to establish the conditions under which the standard condition of cycles, $f'(z_s) < -1$, would hold under the general bilateral trading mechanism.

We can show the existence of an endogenous cycle without relying on the standard treatment of $f'(\cdot) < -1$.  To establish a sufficient condition for an endogenous cycle, consider a two-period cycle with $z_1 < z_s < p^*\leq z_2 $. Since $z_2 \geq p^*$, this cycle satisfies $z_1 = \frac{z_2}{1 + i} < z_s < p^*$, where $z_1$ solves
$$L(z_1)= \frac{(1+i)^2-1}{\alpha(1-\sigma+\sigma\chi)}\chi.$$ 
It is straightforward to show that $z_1 < z_s$ because $L'(\cdot)<0$, and $z_s$ solves (\ref{eq:mss}). By checking the condition $z_2=z_1(1+i)>p^*$, we can derive the condition under which the economy exhibits a two-period cycle that satisfies $z_1<z_s<p^*\leq z_2$.  

\begin{figure}[t]
\centerfloat 
\includegraphics[width=8cm,height=7cm]{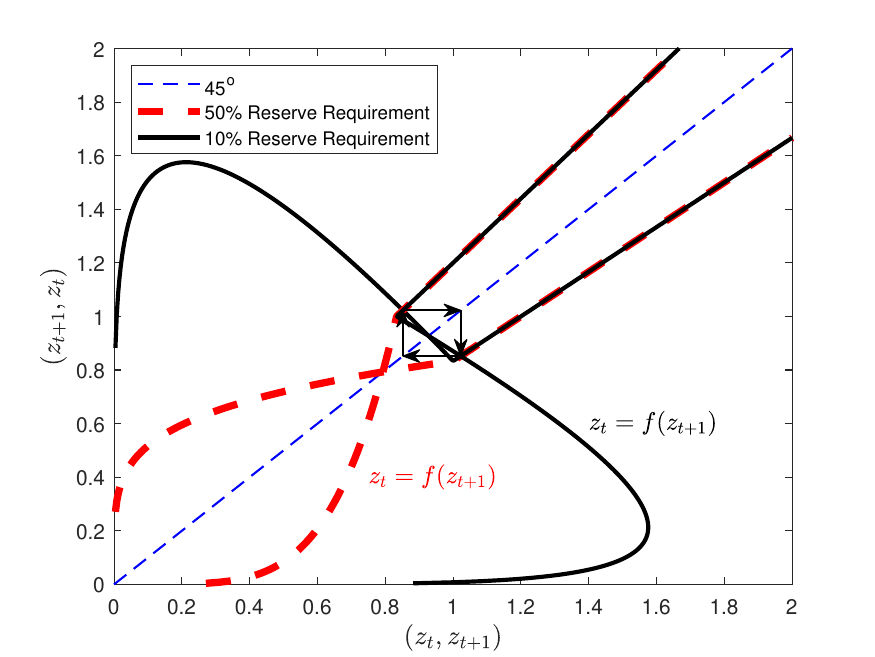}
\includegraphics[width=8cm,height=7cm]{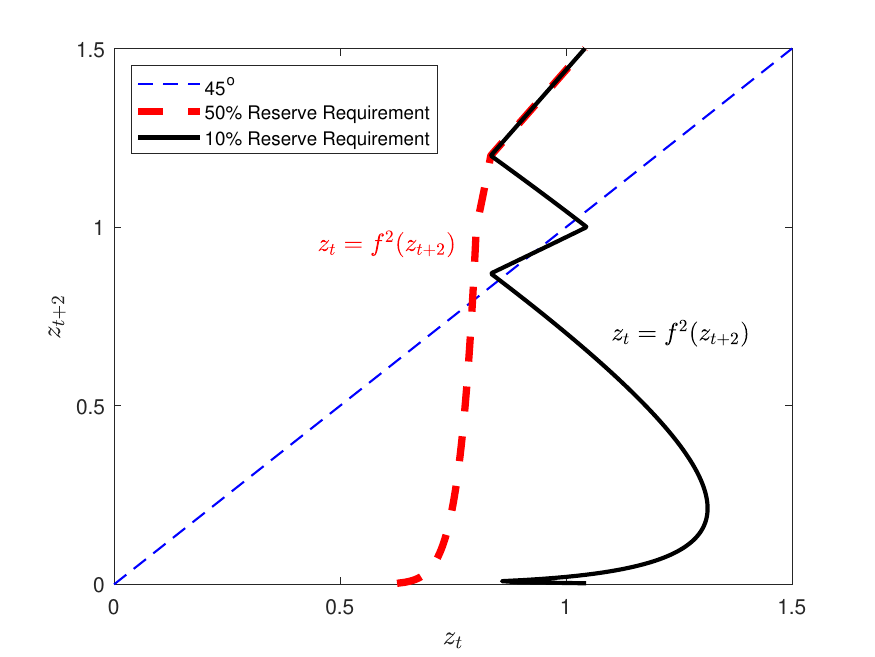}
\caption{A Two-period Cycle under Fractional Reserve Banking}
\label{fig:cycle2}
\end{figure} 

\begin{proposition}[\textbf{Two-period Monetary Cycle}]
\label{prop:2cycle}
There exists a two-period cycle with $z_1<z_s < p^*\leq z_2$ if $\chi\in(0,\bar{\chi}_m]$, where
$$\bar{\chi}_m\equiv\frac{(1-\sigma)\alpha L\left(\frac{p^*}{1+i}\right)}{(1+i)^2-1-\sigma\alpha L\left(\frac{p^*}{1+i}\right)}.$$
When this type of two-period cycle exists, lowering $\chi$ increases 
the difference
between peak and trough, $z_2-z_1$.
\end{proposition}
\begin{proof}
See Appendix \ref{sec:proof}.
\end{proof}

Proposition \ref{prop:2cycle} shows that, under the general trading mechanism, lowering the reserve requirement can induce a two-period cycle and increase the volatility of the real balances. By lowering $\chi$, the liquidity premium dominates the storage value. Consequently, $f(\cdot)$ is more likely to exhibit the backward bending feature, which can lead to an endogenous cycle. Figure \ref{fig:cycle2} shows an example of this case.

Now, let's introduce some additional assumptions to determine the condition for $\chi$ such that $f'(z_s) < -1$. Consider a special case where $-qu''(q)/u'(q)=\eta$, $c(q)=q$, and the buyer makes a take-it-or-leave-it (TIOLI) offer. In this case, as $L(z_s)=u'(q)-1$ and $L'(z_s)z_s=z_s u''(z_s)=-\eta u'(z_s)$, we can rewrite (\ref{eq:fprime1}) as follows:
\begin{equation}\label{eq:fprime2}
f'(z_s)=\frac{1}{1+i}\left\{\frac{1-\sigma+\sigma\chi}{\chi}\alpha\left[
u'(z_s)(1-\eta)-1\right]+1\right\}<-1
\end{equation}
where $u'(z_s)=1+\frac{i\chi}{\alpha(1-\sigma+\sigma\chi)}$. Solving (\ref{eq:fprime2}) for $\chi$ yields the following proposition.

\begin{proposition}
\label{prop:speical}
Assume $-qu''(q)/u'(q)=\eta$, $c(q)=q$, and the buyer makes take-it-or-leave-it offer to the seller. If  $\chi\in(0,\chi_m)$, where   
\begin{equation}
\label{eq:cycle2}
\chi_m\equiv\frac{\alpha\eta(1-\sigma)}{\eta(1-\alpha\sigma)+(2-\eta)(1+i)},
\end{equation}
then $f'(z_s)<-1$.
\end{proposition}
\begin{proof}
See Appendix \ref{sec:proof}.
\end{proof} 
Since $\chi<\chi_m$ implies $f'(z_s)<-1$, 
following the standard textbook method (see \citealp*{azariadisintertemporal}), we can show that if $\chi<\chi_m$, there exists a two-period cycle with $z_1<z_s<z_2$. Whereas (\ref{eq:cycle2}) is written in terms of $\chi$, this condition can be written in terms of $i$, as follows:
\begin{equation}
\label{eq:othercond}
0<i<\frac{\eta[\alpha(1-\sigma)-\chi(1-\alpha\sigma)]}{\chi(2-\eta)}
\end{equation}
The role of $i$ on cycles depends on $\eta$. By (\ref{eq:othercond}), if $\eta<2$, lowering either $\chi$ or $i$ can induce a cycle. If $2/(\alpha\sigma)>\eta>2$, $\chi_m$ is negative when $i>\frac{\eta\alpha\sigma-2}{2-\eta}$ and positive when $i<\frac{\eta\alpha\sigma-2}{2-\eta}$. In this case, setting $i$ higher than $\frac{\eta\alpha\sigma-2}{2-\eta}$ eliminates cyclic equilibria. If $\eta\geq 2/(\alpha\sigma)$,  $\chi_m$ is negative for all $i$, implying the cycle does not exist.  When $\eta=2$, $\chi_m$ is constant, implying that the $i$ has no effect on the cycle in this case. 

\begin{figure}[tp!]
\centerfloat 
\includegraphics[width=8cm,height=7cm]{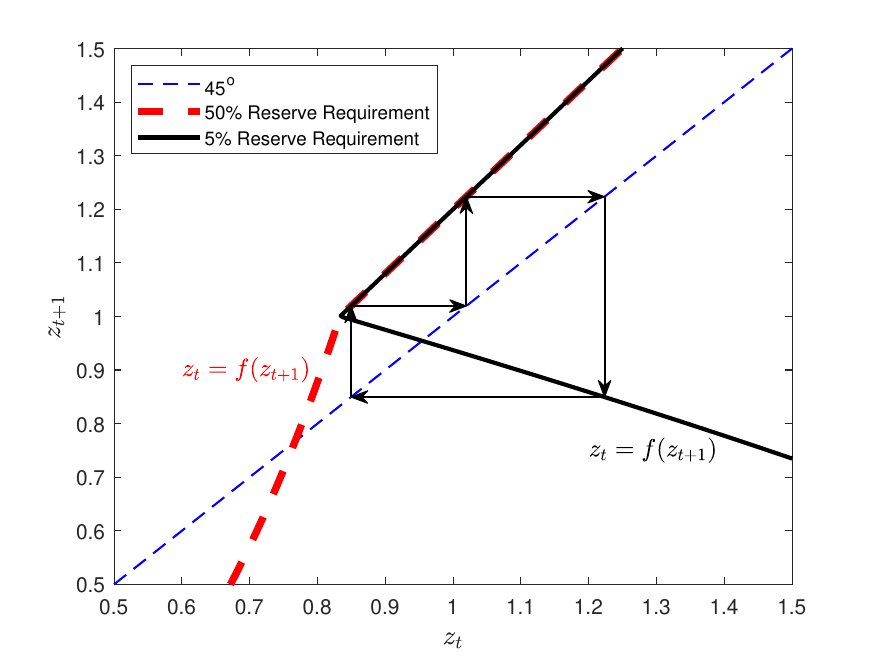}
\includegraphics[width=8cm,height=7cm]{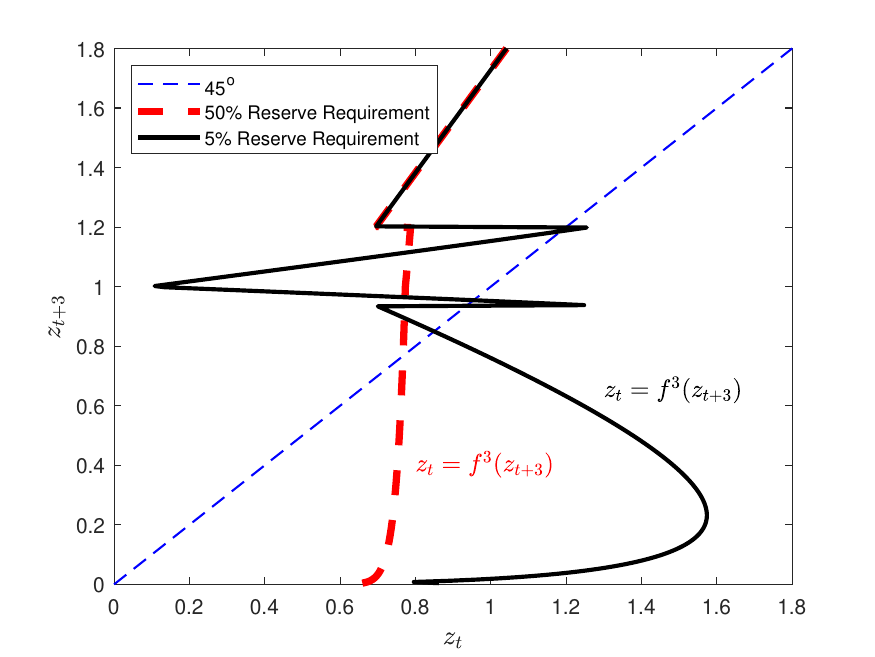}
\caption{A Three-period Cycle under Fractional Reserve Banking}
\end{figure} 

In addition to the conditions for a two-period cycle, the next result provides the condition for a three-period cycle under the general trading mechanism. The existence of three period-cycles implies cycles of all orders as well as chaotic dynamics (see \citealp*{sharkovskii1964cycles} and \citealp*{li1975period}).
\begin{proposition}[\textbf{Three-period Monetary Cycle and Chaos}]
\label{prop:3cycle}
A three-period cycle with $z_1<z_2<p^*\leq z_3$ does not exist. There exists a three-period cycle with $z_1<p^*\leq z_2<z_3$ if $\chi\in(0,\hat{\chi}_m]$, where
$$\hat{\chi}_m\equiv\frac{(1-\sigma)\alpha L\left(\frac{p^*}{1+i}\right)}{(1+i)^3-1-\sigma\alpha L\left(\frac{p^*}{1+i}\right)}.$$
When this type of three-period cycle exists, lowering $\chi$ increases 
the difference
between peak and trough, $z_3-z_1$.
\end{proposition}
\begin{proof}
See Appendix \ref{sec:proof}.
\end{proof}

The following corollary is a direct result from Proposition \ref{prop:3cycle}. 

\begin{corollary}[\textbf{Binding Liquidity Constraint}]
\label{coro:least}
In any n-period cycle, the liquidity constraint binds, $z_t<p^*$, at least one periodic point over the cycle.
\end{corollary}
\begin{proof}
See Appendix \ref{sec:proof}.
\end{proof}

The model can also generate sunspot cycles. Consider a Markov sunspot variable $S_t\in\{1,2\}$. This sunspot variable is not related to fundamentals but may affect equilibrium. Let $\Pr(S_{t+1}=1|S_{t}=1)=\zeta_1$ and $\Pr(S_{t+1}=2|S_{t}=2)=\zeta_2$. The sunspot is realized in the FM. Let $W^{S}_{t}$ be the CM value function in state $S$ in period $t$, then 
\begin{equation*}
\begin{split}
W^{S}_{t}(m_{t},d_{t},\ell_{t})=&\max_{X_{t},H_{t},\hat{m}_{t+1}} U(X_t)-H_t+\beta\left[\zeta_s G^{S}_{t+1}(\hat{m}_{t+1})+ (1-\zeta_s)G^{-S}_{t+1}(\hat{m}_{t+1})\right] \\
\text{s.t. } & \phi^{S}_{t}\hat{m}_{t+1}+X_{t}=H_{t}+T_{t}+\phi^{S}_{t} m_{t}+(1+i_{d,t})\phi^{S}_{t} d_{t}-(1+i_{l,t})\phi^{S}_{t}\ell_{t}.
\end{split}
\end{equation*}
The FOC can be written as
\begin{equation}
\label{eq:sunspot1}
-\phi^{S}_{t}+\beta\zeta_s G'^{S}_{t+1}(\hat{m}_{t+1})+\beta(1-\zeta_s)G'^{-S}_{t+1}(\hat{m}_{t+1})=0.
\end{equation}
Solving the FM problem results in
\begin{equation}
\label{eq:sunspot2}
G'^{S}_{t+1}(m^{S}_{t+1})=\phi^{S}_{t+1}\left[ \frac{1-\sigma+\sigma\chi}{\chi}\alpha L(z^{S}_{t+1})+1 \right]. 
\end{equation}
We substitute (\ref{eq:sunspot2}) into (\ref{eq:sunspot1}) and use the money market clearing condition $m_{t+1}$ = $M_{t+1}$ to get the Euler equation.  
\begin{align}
\phi^{S}_{t}&=\beta\zeta_s \phi^{S}_{t+1}\left[\frac{1-\sigma+\sigma\chi}{\chi}\alpha L(z^{S}_{t+1})+1\right] +\beta(1-\zeta_s)  \phi^{-S}_{t+1}\left[\frac{1-\sigma+\sigma\chi}{\chi}\alpha L(z^{-S}_{t+1})+1\right]. \nonumber \end{align} where $z^S_{t+1}=\phi^S_{t+1} M_{t+1}(1-\sigma+\sigma\chi)/\sigma\chi$.
Then multiply both sides of the Euler equation by $M_{t}(1-\sigma+\sigma\chi)/\sigma\chi$ to reduce the equilibrium condition into one difference equation of real balances $z^S_{t+1}$:
\begin{align}
\label{eq:sunspot3}
z^{S}_{t}
&=\frac{\zeta_s z^{S}_{t+1}}{1+i}\left[\frac{1-\sigma+\sigma\chi}{\chi}\alpha L(z^{S}_{t+1})+1\right]+\frac{(1-\zeta_s) z^{-S}_{t+1}}{1+i}\left[\frac{1-\sigma+\sigma\chi}{\chi}\alpha L(z^{-S}_{t+1})+1\right] \nonumber  \\
&=\zeta_s f(z^{S}_{t+1})+(1-\zeta_s) f(z^{-S}_{t+1}).
\end{align}
We define a sunspot equilibrium as follows:

\begin{definition}[\textbf{Proper Sunspot Equilibrium}]
A proper sunspot equilibrium consists of the sequences of real balances $\{z^S_{t}\}_{t=0, S=1,2}^{\infty}$ and probabilities $(\zeta_1, \zeta_2)$, solving (\ref{eq:sunspot3}) for all $t$.
\end{definition} 

Consider stationary sunspot equilibria with $z^1<z^2$ that only depend on the state, not the time. The liquidity constraint is binding in state $S=1$. By the standard approach (see again \citealp*{azariadisintertemporal} for the textbook treatment), the condition for two-period cycles is also sufficient and necessary for two-state sunspot equilibrium. If $f'(z_s)<-1$ or $\chi<\bar{\chi}_m$, there exists $(\zeta_1,\zeta_2)\in(0,1)^2$, $\zeta_1+\zeta_2<1$, such that the economy has a proper sunspot equilibrium in the neighborhood of $z_s$.

\begin{proposition}[\textbf{Stationary Sunspot Equilibrium}]\label{prop:sunspot}
The stationary sunspot equilibrium exists if either $\chi<\chi_m$ or $f'(z_s)<-1$.
\end{proposition}

\begin{proof}
See Appendix \ref{sec:proof}.
\end{proof}

In addition to the deterministic and stochastic cycles, the model also features the equilibria where real balance increases above the steady-state until certain time, $T$, and crashes to zero. Consider a sequence of real balances $\{z_t\}_{t=0}^\infty$  with $z_T\equiv\max \{z_t\}^{\infty}_{t=0}>z_s$ (bubble) that crashes to 0 (burst) as $t\rightarrow\infty$, where $T\geq 1$ and $z_T>z_0$. We refer to this equilibrium as a self-fulfilling bubble and burst equilibrium:

\begin{figure}[tp!]
\centerfloat 
\includegraphics[width=8cm,height=7cm]{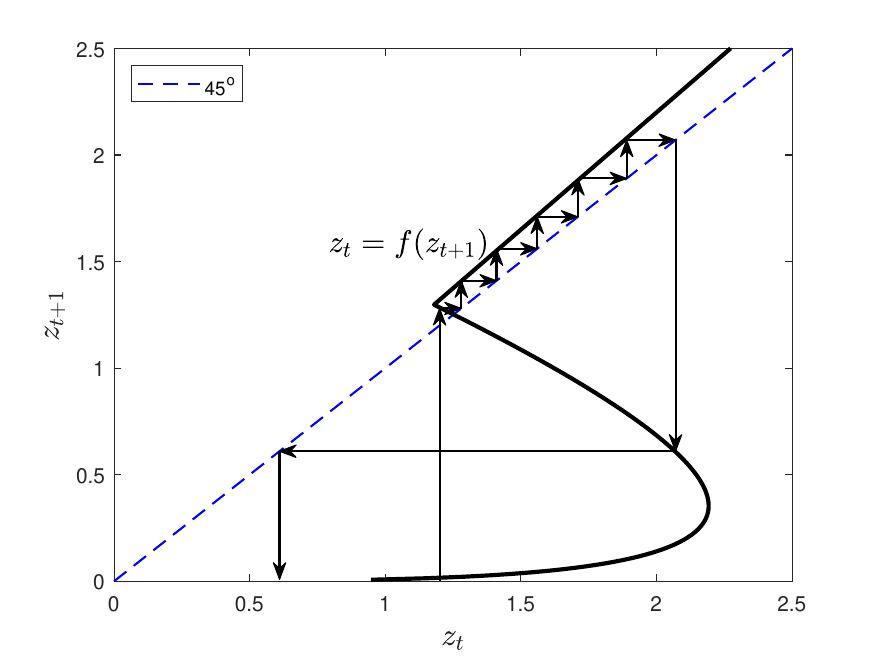}
\includegraphics[width=8cm,height=7cm]{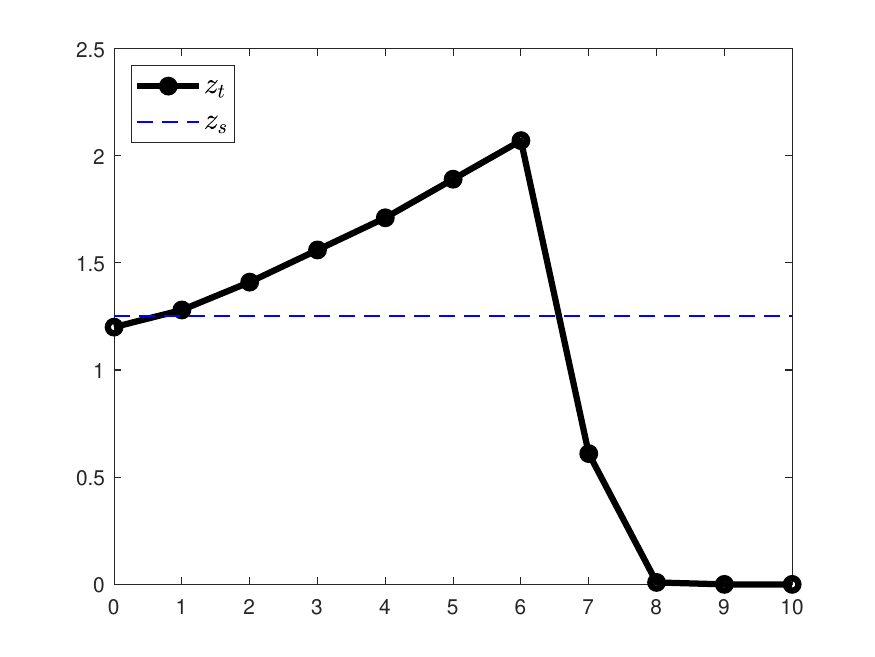}
\caption{Self-Fulfilling  Bubble  and  Burst  Equilibrium}
\label{fig:bubble_burst}
\end{figure}

\begin{definition}[\textbf{Self-Fulfilling Bubble and Burst Equilibrium}]
For initial real balance $z_0>0$, a self-fulfilling bubble and burst equilibrium is a sequence of   $\{z_t\}^{\infty}_{t=0}$ satisfying (\ref{eq:mdynamics})
and $0<z_s<z_{T}$, 
$\lim_{t\rightarrow\infty}z_t=0$ where $z_T=\max \{z_t\}^{\infty}_{t=0}$ with $T\geq 1$.
\end{definition}

Figure \ref{fig:bubble_burst} illustrates an example. In Figure \ref{fig:bubble_burst}, $f$ is not monotone, so $f^{-1}$ is a correspondence. When $f$ is not monotone, there are multiple equilibrium paths for $\{z_{t+1}\}$ over some range for $z_t$. This example starts at $z_0$, which is lower than $z_s$, and then increases, surpassing $z_s$, repeatedly rising until it reaches $z_6$. Afterward, it crashes and eventually converges to $0$. During the bubble, the return on money is equal to $1/(1+i)$, and liquidity is abundant. However, the real balances cannot continue to increase indefinitely otherwise it would violate the transversality condition. The real balance increases until it reaches a certain point, after which the economy crashes and moves toward a non-monetary equilibrium. The timing of these crashes is indeterminate.

The next step is to examine the conditions under which this type of equilibrium can occur. When $z_s>\bar z$, where $\bar z$ satisfies $f'(\bar z)=0$, multiple equilibria exist. This implies that solving $z_t=f(z_{t+1})$ given $z_t$ yields multiple solutions for $z_{t+1}$. If $f(\bar z)\geq q^*$, the self-fulfilling bubble and burst equilibrium exists. Assuming $-qu''(q)/u'(q)=\eta$, $c(q)=q$, and a buyer makes a TIOLI offer to the seller, Proposition \ref{prop:burst} shows that lowering the reserve requirement can induce this type of equilibrium.

\begin{proposition}[\textbf{Existence of Self-Fulfilling Bubble and Burst Equilibrium}]
\label{prop:burst}
Assume $-qu''(q)/u'(q)=\eta$, $c(q)=q$, and the buyer makes take-it-or-leave-it offer to the seller. There exist a self-fulfilling bubble and burst equilibrium, if
$$0<\chi<\min\left\{\frac{(1-\sigma)\alpha\eta(1+i)}{(1-\eta)^2 q^*+(1+i)[(1-\eta)(3+i-\eta)-\alpha\sigma\eta]}, \frac{\alpha\eta(1-\sigma)}{1+i-\eta(i+\alpha\sigma)}\right\} $$
\end{proposition}
\begin{proof}
See Appendix \ref{sec:proof}.
\end{proof}

\section{Money and Unsecured Credit}
\label{sec:credit}

Consider an alternative payment instrument in the DM - unsecured credit. The buyer can pay for DM goods using unsecured credit that will be redeemed to the seller in the following CM and she can borrow up to her debt limit, $\bar{b}_t$. For simplicity, I assume that the buyer makes a TIOLI offer to the seller in the DM, which means the buyer maximizes her surplus subject to the seller’s participation constraint. The DM cost function is $c(q)=q$.  Suppose the buyer has issued $b_t$ units of unsecured debt in the previous DM (or, if $b_t<0$, the seller has extended unsecured loans to the buyer from the previous DM). The CM value function is 
\begin{equation}
\begin{split}
W_{t}(a_{t},s_{t},\ell_{t},-b_t)=&\max_{X_{t},H_{t},\hat{m}_{t+1}} U(X_t)-H_t+\beta G_{t+1}(\hat{m}_{t+1}) \\
\text{s.t. } \phi_{t}\hat{m}_{t+1}+X_{t}=H_{t}+T_{t}& +\phi_{t} a_{t}+(1+i_{s,t})\phi_{t} s_{t}-(1+i_{l,t})\phi_{t}\ell_{t}-b_t,
\end{split}
\end{equation}
which is the same as before except that the agent needs to pay or collect the debt. The agent's FM problem is identical to the previous section. Then, $1-\sigma$ fraction of agents will deposit $\hat{m}_{t+1}$, and $\sigma$  fraction of agents will borrow loan from the bank. The DM value function is 
$$ V^b_{t}(m_{t}+d_{t},0,\ell_{t})=\alpha [u(q_{t})-q_{t}]+W_{t}(m_{t}+d_{t},0,\ell_{t},0),$$
where $q_t=\min\{q^*,\bar{b}_t+ (m_t+d_t)\phi_t\}$ and $d_t=\ell_t$. Given $\bar{b}_t$, solving equilibrium yields
\begin{align}
z_{t}&=
\begin{dcases} 
\frac{z_{t+1}}{1+i}\left\{\frac{1-\sigma+\sigma\chi}{\chi}\alpha\left[u'(z_{t+1}+\bar{b}_{t+1})-1\right]+1\right\} &\text{ if } z_{t+1}+\bar{b}_{t+1}< q^* \\
\frac{z_{t+1}}{1+i}             &\text{ if } z_{t+1}+\bar{b}_{t+1}\geq q^*,
\end{dcases}\label{eq:mcdynamics1}
\end{align}
where $z_{t+1}=(1-\sigma+\sigma\chi)\phi_{t+1}M_{t+1}/(\sigma\chi)$. 

Next, I am going to endogenize the debt limit. The buyer cannot commit to pay back the debt. If the buyer reneges  she is captured with probability $\mu$. The punishment for a defaulter is permanent exclusion from the DM trade but she can still produce for herself in the CM. The value of autarky is $\underbar{W}(0,0,0,0)= [U(X^*)-X^*+T]/(1-\beta)$. The incentive condition for voluntary repayment is
\vspace{-0.3cm}
$$\underbrace{-b_t+W_t(a_{t},d_{t},\ell_{t},0)}_{\text{value of honoring debts}}\geq \underbrace{ (1-\mu)W_t(a_{t},d_{t},\ell_{t},0)+\mu\underbar{W}(a_{t},d_{t},\ell_{t},0)}_{\text{value of not honoring debts}}. $$

One can write the debt limit $\bar{b}_t$ as $b_t\leq\bar{b}_t\equiv\mu W_t(0,0,0,0)-\mu\underbar{W}(0,0,0,0)$.
Recall the CM value function. Using the solution of FM, we can rewrite the buyer's CM value function as
\begin{equation*}
\begin{split}
W_t(0,0,0,0)&= U(X^*)-X^*+T_t+\beta W_{t+1}(0,0,0,0) \\
& +\max_{\hat{m}_{t+1}} \{-\phi_{t} \hat{m}_{t+1} +\beta \alpha\sigma [u(q_{t+1})-q_{t+1}]+\beta \phi_{t+1}\hat{m}_{t+1}\},
\end{split}
\end{equation*}
where $q_{t+1}=\min\{q^*,z_{t+1}+\bar{b}_{t+1}\}$.
Substituting $W_t(0,0,0,0)=\bar{b}_t/\mu +\underbar{W}(0,0,0,0)$ and $\hat{m}_{t+1}=M_{t+1}$ yields $$\frac{\bar{b}_t}{\mu}=-\phi_{t} M_{t+1}+\beta \alpha\sigma [u(z_{t+1}+\bar{b}_{t+1})-z_{t+1}-\bar{b}_{t+1}]+
\frac{\beta\bar{b}_{t+1}}{\mu}+\beta \phi_{t+1} M_{t+1},$$
where $M_{t+1}$ and $z_{t+1}$ solve (\ref{eq:mcdynamics1}). Rearranging terms yields
\begin{align}
\bar{b}_t &=
\begin{dcases} 
\beta\bar{b}_{t+1}+
\frac{\chi\mu\sigma[-\gamma z_{t}+\beta z_{t+1}]}{1-\sigma+\sigma\chi} +\beta\alpha\mu\sigma S(z_{t+1}+\bar{b}_{t+1}) &\text{ if } z_{t+1}+\bar{b}_{t+1}< q^* \\
\beta\bar{b}_{t+1}+
\frac{\chi\mu\sigma[-\gamma z_{t}+\beta z_{t+1}]}{1-\sigma+\sigma\chi} +\beta\alpha\mu\sigma S(q^*) &\text{ if } z_{t+1}+\bar{b}_{t+1}\geq q^*,
\end{dcases} \label{eq:mcdynamics2} 
\end{align} where $S(z_{t+1}+\bar{b}_{t+1})\equiv [u(z_{t+1}+\bar{b}_{t+1})-z_{t+1}-\bar{b}_{t+1}]$ is the buyer's trade surplus. The equilibrium can be collapsed into a dynamic system satisfying (\ref{eq:mcdynamics1})-(\ref{eq:mcdynamics2}). 

In the stationary equilibrium,  (\ref{eq:mcdynamics1}) becomes 
\begin{equation}\label{eq:mc_sme1} 
-\frac{i\chi}{\alpha(1-\sigma+\sigma\chi)}+u'(q)-1\leq0, = \text{ if } z>0 
\end{equation}
and (\ref{eq:mcdynamics2}) becomes 
\begin{align}
(1-\beta)\bar{b} &=
\begin{dcases} 
\frac{\chi\mu\sigma[\beta-\gamma]z}{1-\sigma+\sigma\chi} +\beta\alpha\mu\sigma [u(z+\bar{b})-z-\bar{b}] &\text{ if } z+\bar{b}< q^* \\
\frac{\chi\mu\sigma[\beta-\gamma]z}{1-\sigma+\sigma\chi} +\beta\alpha\mu\sigma [u(q^*)-q^*] &\text{ if } z+\bar{b}\geq q^*,
\end{dcases} \label{eq:mc_sme2} 
\end{align} 
where $q=\min\{z+\bar{b},q^*\}$. The stationary equilibrium solves the above two equations, and it falls into one of the three cases: the pure money equilibrium, the pure credit equilibrium, and the money-credit equilibrium. First, if no one can capture the buyer after she reneges, $\mu=0$, the unsecured credit is not feasible, $\bar{b}=0$. In this case, the equilibrium will be the pure money equilibrium. Second, when $\bar{b}$ solving (\ref{eq:mc_sme2}) satisfies $u'(\bar{b})<i\chi/[\alpha(1-\sigma+\sigma\chi)]$ then money is not valued, $z=0$. We have the pure credit equilibrium in this case. Third, if solutions of (\ref{eq:mc_sme1})-(\ref{eq:mc_sme2}), $(z,\bar{b})$ are strictly positive then we have the money-credit equilibrium.

The debt limit at the stationary equilibrium, $\bar{b}$, is a fixed point satisfying $\bar{b}=\Omega(\bar{b})$ where
\begin{align}
\Omega(\bar{b})=
\begin{dcases} 
\frac{\mu\sigma\alpha}{\rho}[u(\tilde{q})-\tilde{q}] -\frac{i\mu\sigma \chi}{\rho(1-\sigma+\sigma\chi)}(\tilde{q}-\bar{b})     &\text{ if } \tilde{q}>\bar{b}\geq 0\\
\frac{\mu\sigma\alpha}{\rho}[u(\bar{b})-\bar{b}] &\text{ if } q^*>\bar{b}\geq \tilde{q} \\
\frac{\mu\sigma\alpha}{\rho}[u(q^*)-q^*] &\text{ if } \bar{b}\geq q^*
\end{dcases}  
\end{align}
where $\tilde{q}$ solves $u'(\tilde{q})=1+i\chi/[\alpha(1-\sigma+\sigma\chi)]$ and $\rho\equiv 1/\beta-1$.
The DM consumption $q_s$ is determined by $q_s=\min\{q^*,\max\{\tilde{q},\bar{b}\}\}$. Money and credit coexist if and only if $0<\bar{b}<\tilde{q}$, which holds when $0<\mu<\min\{1,\tilde{\mu}\}$,
where 
$$\tilde{\mu}\equiv\frac{\rho(1-\sigma+\sigma\chi)}{\alpha\sigma[u(\tilde{q})/\tilde{q}-1](1-\sigma+\sigma\chi)-2i\sigma\chi}.$$
The DM consumption is decreasing in $i$ in the stationary monetary equilibrium. 


Consider the dynamics of equilibria where money and credit coexist. I claim the main results from Section \ref{sec:results} - lowering the reserve requirement can induce endogenous cycles - still hold even after unsecured credit is introduced. It is clear that the standard treatment $f'(z_s)<-1$ from \cite{azariadisintertemporal} cannot be used here because now the equilibrium consists of a system of equations. Instead, I apply the approach used in Proposition \ref{prop:2cycle} and \ref{prop:3cycle}. 
For compact notation, let $\iota\equiv\max\{i,\rho\}$ and  $w_j\equiv z_j+\bar{b}_j$. The following proposition establishes the conditions for a two-period cycle, a three-period cycle, and chaotic dynamics.

\begin{proposition}[\textbf{Monetary Cycles with Unsecured Credit}]
\label{prop:mc3cycle}
There exists a two-period cycle of money and credit with $w_1<q^*<w_2$ if $\chi\in(0,\bar\chi_c)$, where 
$$\bar\chi_c\equiv\frac{(1-\sigma)\alpha\left[u'\left(\frac{q^*}{1+\iota}\right)-1\right]}{(1+i)^2-1-\sigma\alpha\left[u'\left(\frac{q^*}{1+\iota}\right)-1\right]}.$$
There exists a three-period cycle of money and credit with $w_1<q^*<w_2<w_3$, if $\chi\in(0,\hat{\chi}_c)$, where
$$\hat{\chi}_c\equiv\frac{(1-\sigma)\alpha\left[u'\left(\frac{q^*}{1+\iota}\right)-1\right]}{(1+i)^3-1-\sigma\alpha\left[u'\left(\frac{q^*}{1+\iota}\right)-1\right]}.$$
\end{proposition}
\begin{proof}
See Appendix \ref{sec:proof}.
\end{proof}

\section{Calibrated Examples}
\label{sec:quant}

\subsection{Parameters and Targets}

In this section, I calibrate the model using U.S. data. I calibrate both the model without unsecured credit and the model with unsecured credit. First, assuming there is no unsecured credit, I calibrate the model without unsecured credit, which will be referred to as Model 1 throughout this section. In addition to Model 1, I also calibrate the model with unsecured credit, referred to as Model 2. For monetary aggregates, I use M1 adjusted for retail sweep accounts as in \cite{aruoba2011money} and \cite{venkateswaran2014pledgability}.
Following to \cite{krueger2006does},  revolving consumer credit series are used as unsecured credit.

I set the discount rate $\beta=0.9709$  to match the real annual interest rate of $3\%$. Using the average required reserve to deposit ratio for 1980Q1-2008Q4, the benchmark required reserve ratio is set to $7.77\%$.\footnote{To compute the required reserve ratio, I divide the required reserve ratio, I divide the required reserves by the deposit component of sweep-adjusted M1. The average of this ratio for 1980Q1-2008Q4 is 0.0777} The benchmark value for $i$ is set to $0.0564$ as the average annualized nominal interest rate is $5.64\%$.  The matching function in the DM is  $\mathcal{M}(\mathcal{B},\mathcal{S})=\frac{\mathcal{B}\mathcal{S}}{\mathcal{B}+\mathcal{S}}$,  where $\mathcal{B}$ and $\mathcal{S}$ denotes the measure of buyers and sellers, respectively. This implies $\alpha=\mathcal{M}(\sigma,1-\sigma)/\sigma=(1-\sigma)$ and $\alpha_s=\mathcal{M}(\sigma,1-\sigma)/(1-\sigma)=\sigma$. The fraction of buyer is set to $\sigma=0.5$ for a normalization. The utility functions for the parameterization are  
$$U(X)=B\log(X),\qquad u(q)=\frac{Cq^{1-\eta}}{1-\eta}$$ implying  $X^*=B$ and the DM cost function is given as $c(q)=q$.  Assume the buyer makes a take-it-or-leave-it offer to the seller in the DM trade, implying $\lambda(q)=Cq^{-\eta}-1$.

This section focuses on the equilibrium where $X>0$ which requires $U(X)-X>0$. To guarantee $U(X)-X>0$, we need to have $\log(B)>1$ since $B=X^*$. Otherwise, the CM consumption is zero, $X=0$. For normalization, I set $B=3$. The parameters $(C,\eta)$ are set to match the money demand relationship. In the model, the money demand relationship is given by real balances of money as a fraction of output
$$Z\equiv\frac{z}{y}=\frac{z}{B+\sigma\alpha q}$$
where $y$ is the real output of the economy and elasticity of $z/y$ with respect to $i$
$$\frac{\partial\log(Z)}{\partial \log(i)}=\frac{i}{Z}\frac{\partial Z}{\partial i}.$$

\begin{table}[tp!]
\footnotesize
\caption{Model Parametrization}
\vspace{-0.2cm}
\centering
\begin{threeparttable}
\begin{tabular}{lrrr}
\toprule \hline
& Data    &  \multicolumn{1}{r}{Model 1}  & \multicolumn{1}{r}{Model 2}    \\ 
&  &  No Credit, $\mu=0$ & With Credit, $\mu>0$   \\ \hline
\textbf{Parameters}    & & & \\
DM utility level, $C$         &   & 0.9107 &  1.0452 \\ 
DM utility curvature, $\eta$      &    &   0.1409 & 0.3222 \\
Monitoring probability, $\mu$      &    &   - &  0.1359 \\
\textbf{Targets}    & & & \\
avg. $z/y$              & 0.1473 & 0.1475  &    0.1474          \\  
elasticity of $z/y$ wrt $i$ & -0.0661 & -0.0661
 & -0.0662            \\   
avg. $\sigma\alpha b/y$ & 0.0466 & -
 & 0.0465         \\    
\hline \bottomrule
\end{tabular}

\end{threeparttable}
\label{tab:param}
\end{table}

Specifically, the parameter $C$ is set to match the average money stock to GDP ratio, which is $0.1473$, and the parameter $\eta$ is set to match the elasticity of $z/y$ with respect to $i$. The target elasticity is estimated using the following regression:
$$\log(Z_t)=\beta_0 +\beta_1 \log(i_t)+\varepsilon_t $$
The estimated elasticity is $\beta_1 = -0.0661$. In Model 2, agents can use unsecured credit in DM trade, and the monitoring probability $\mu$ is set to match the amount of unsecured credit in the economy as a fraction of output, which is $0.0466$. Its model counterpart is $\sigma\alpha b/y$. Table \ref{tab:param} shows the calibrated parameters and the target moments.

\subsection{Model Implied Thresholds}

\begin{figure}\centerfloat 
  \begin{subfigure}{0.5\textwidth}
    \includegraphics[width=\linewidth]{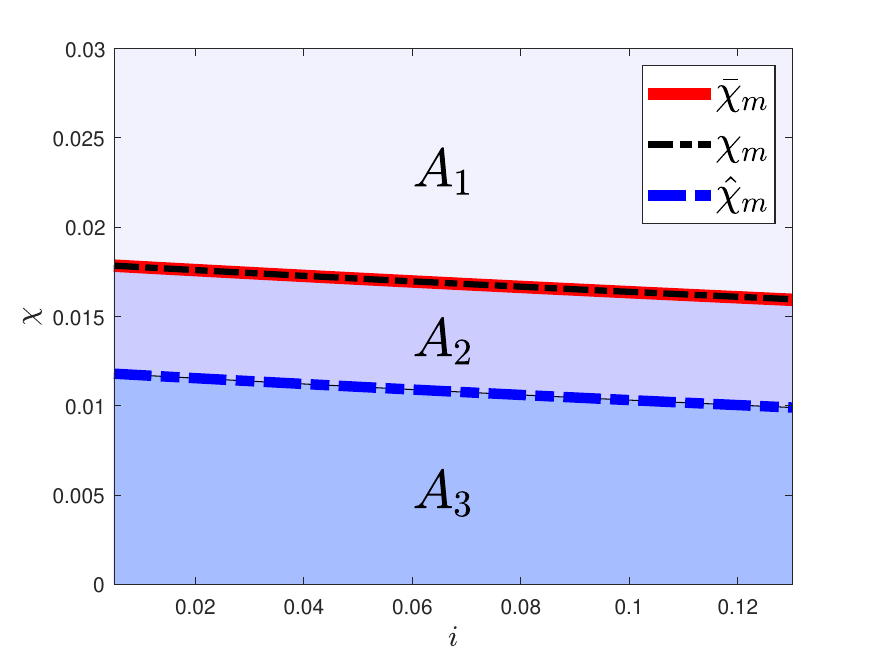}
    \caption{Model 1} \label{fig:thred_1_1}
  \end{subfigure}%
  \begin{subfigure}{0.5\textwidth}
    \includegraphics[width=\linewidth]{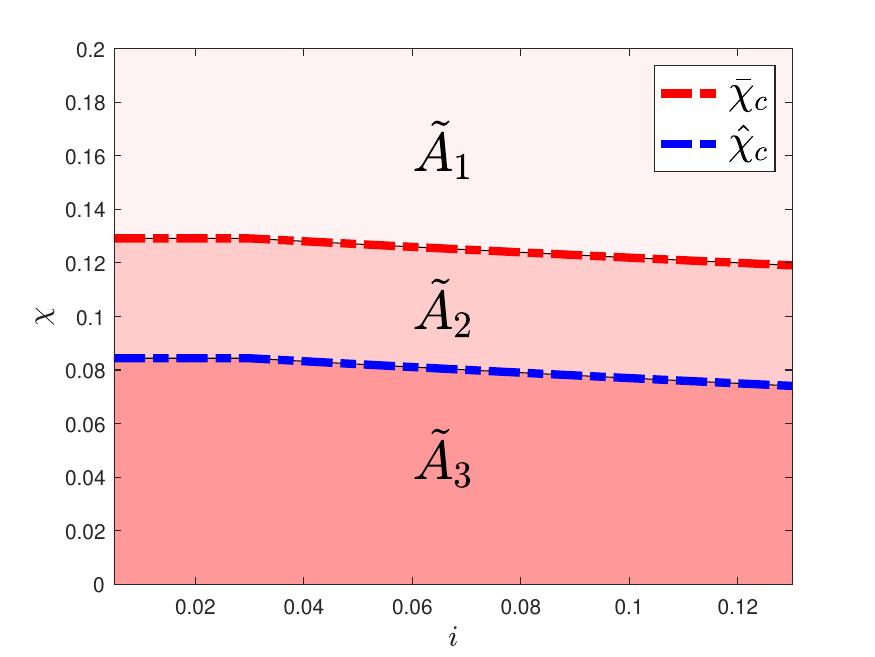}
    \caption{Model 2} \label{fig:thred_1_2}
  \end{subfigure}
\caption{Calibrated Examples I}
\label{fig:thred_1}
\end{figure}

Given the parameterization from the previous section, this section examines whether the conditions for endogenous cycles hold or not. Figure \ref{fig:thred_1_1} displays the thresholds for a two-period cycle, $\chi_m$ and $\bar{\chi}_m$, and the threshold for a three-period cycle and chaotic dynamics, $\hat{\chi}_m$, under different values of $i$ in Model 1. The thresholds decrease as $i$ increases. The thresholds for two-period cycles, $\chi_m$ and $\bar{\chi}_m$, are almost the same and range from 0.0160 to 0.0178. The threshold for a three-period cycle, $\hat{\chi}_m$, varies from 0.0099 to 0.0118. Area $A_1$ denotes the region where $\chi>\chi_m$, indicating the absence of cycles. Area $A_2$ represents the region where $\chi_m>\chi>\hat{\chi}_m$. In area $A_2$, there are two-period cycles, but higher-order cycles may not exist. Area $A_3$ indicates the region where $\chi<\hat{\chi}_m$, implying the presence of higher-order cycles and chaotic dynamics.

Figure \ref{fig:thred_1_2} presents the thresholds for the two-period cycles, $\bar\chi_c$ and three-period cycles and chaotic dynamics, $\hat{\chi}_c$ in Model 2. The threshold $\bar\chi_c$ varies from 0.1192 to 0.1291, and the threshold $\hat{\chi}_c$ ranges from 0.0743 to 0.0844. The thresholds for cycles in Model 2 are higher than those in Model 1 and decrease as $i$ increases. Area $\tilde{A}_1$, $\tilde{A}_2$, and $\tilde{A}_3$ denotes the regions where $\chi>\bar\chi_c$, $\chi_c>\chi>\hat{\chi}_c$, and $\chi<\hat{\chi}_c$, respectively.

\begin{figure}\centerfloat 
  \begin{subfigure}{0.5\textwidth}
    \includegraphics[width=\linewidth]{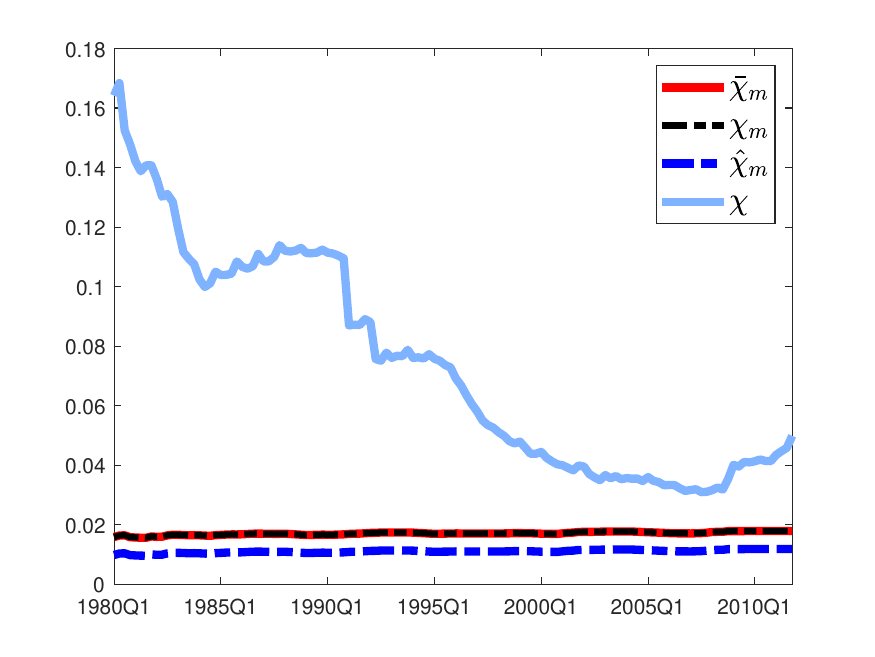}
    \caption{Model 1} \label{fig:thred_2_1}
  \end{subfigure}%
  \begin{subfigure}{0.5\textwidth}
    \includegraphics[width=\linewidth]{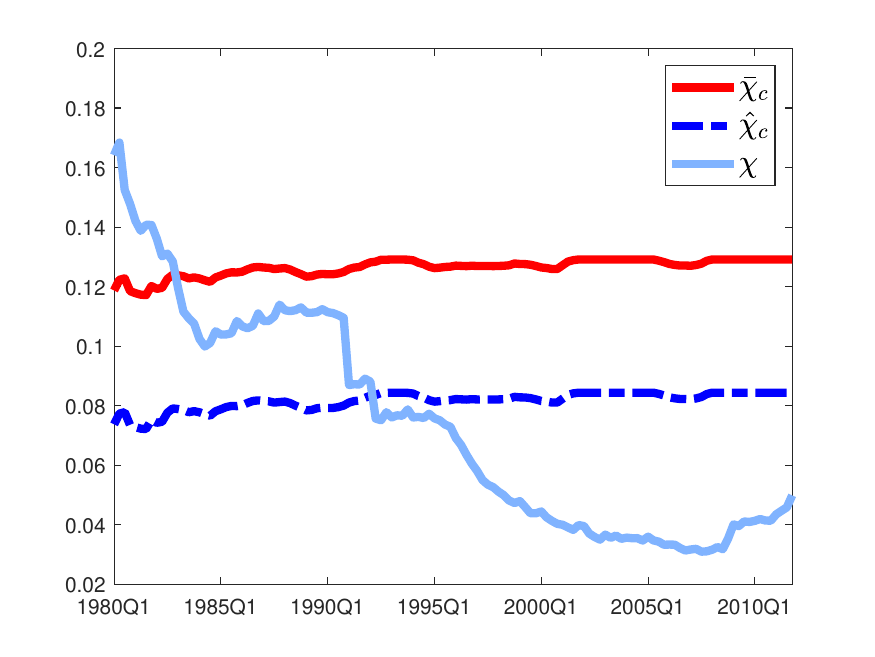}
    \caption{Model 2} \label{fig:thred_2_2}
  \end{subfigure}
\caption{Calibrated Examples II}
\label{fig:thred_2}
\end{figure}

We can compare the thresholds with the actual data.\footnote{To compute $\chi$, I divide the required reserves by deposit component of sweep-adjusted M1.}  Figure \ref{fig:thred_2_1} plots the model-implied thresholds from Model 1 over time. While both of $\chi_m$ and $\hat{\chi}_m$ are higher than $\chi$, they are close to $\chi$ after 2000. Figure \ref{fig:thred_2_2} plots the model-implied thresholds from Model 2 over time. This figure shows that both $\bar\chi_c$ and $\hat{\chi}_c$ are lower than $\chi$ after 1992, and $\bar\chi_c$ has been lower than $\chi$ since 1983. This indicates that the economy could exhibit endogenous fluctuations and chaotic dynamics. It implies that the economy may have endogenous cycles as well as sunspot cycles due to fractional reserve banking, which is independent of the presence of exogenous shocks and changes in fundamentals.

In both models, the endogenous cycle thresholds are not low enough to be ignored, indicating that this channel of volatility needs to be considered in addition to economic fluctuations induced by exogenous shocks that disrupt the dynamic system.

\subsection{News Shocks}
\label{sec:news}

This section explores the role of reserve requirements in the dynamics resulting from news about future changes in monetary policy. To analyze the impact of such news, I follow \cite{gu2019effects} who study the effect of news in the economy where liquidity plays a role. \cite{gu2019effects} show that the response to the announcement can be complicated, and it highly depends on parameters. Using calibrated parameters, this section examines how reducing the reserve requirement can complicate the effects of the monetary policy announcement.

Let the central bank change $i$ permanently, from $i_0$ to $i_T$.\footnote{The permanent change in $i$ implies a permanent change in the rate of monetary expansion, $\gamma$, from $\gamma_0=\beta(1+i_0)$ to $\gamma_T=\beta(1+i_T)$.} Suppose news on changes in the monetary policy at $T$ is announced at time 0.  As in \cite{gu2019effects}, I focused on unique transition consistent with stationarity after information shock. Initially, the economy is in its unique stationary equilibrium for a given $i_0$.  At $t=0$ it is announced that $i$ will change to $i=i_T$ at $t=T$ and stay at $i_T$ permanently. Therefore the stationary equilibrium $z_T$ for a given $i_T$ is a fixed terminal condition that pins down the transition by backward induction. 

First, consider the dynamics without unsecured credit, i.e.,  $\mu=0$.  The permanent change in $i$ from $i_0$ to $i_T$ implies a shift of equation  (\ref{eq:mdynamics}) from $z_{t}=f_0(z_{t+1};i_0,\chi)$ to $z_{t}=f_T(z_{t+1};i_T,\chi)$. Let $z_s(i,\chi)$ be a steady state for a given monetary policy, $(i,\chi)$. Then, the transition starts in steady state with $z_s(i_0,\chi)$ and ends with $z_s(i_T,\chi)$. This can be solved by backward induction, as follows. 
$$z_{T}=f_T(z_{T}),\quad z_{T-1}=f_0(z_{T}),\quad z_{T-2}=f_0(z_{T-1}),\quad...\quad z_{0}=f_0(z_{1})$$

\begin{figure}[tp!]
\centering
\includegraphics[width=7cm,height=5cm]{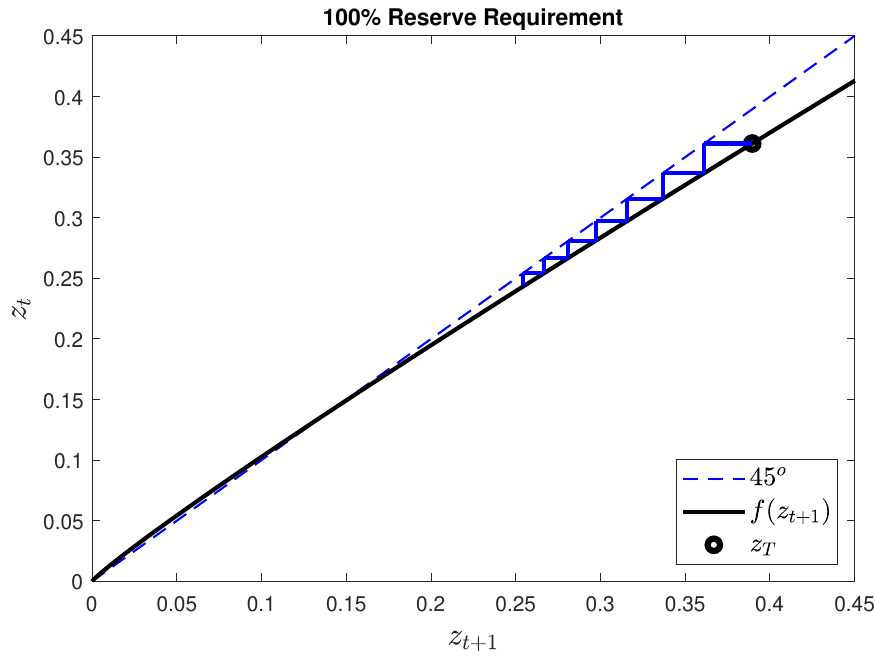}
\includegraphics[width=7cm,height=5cm]{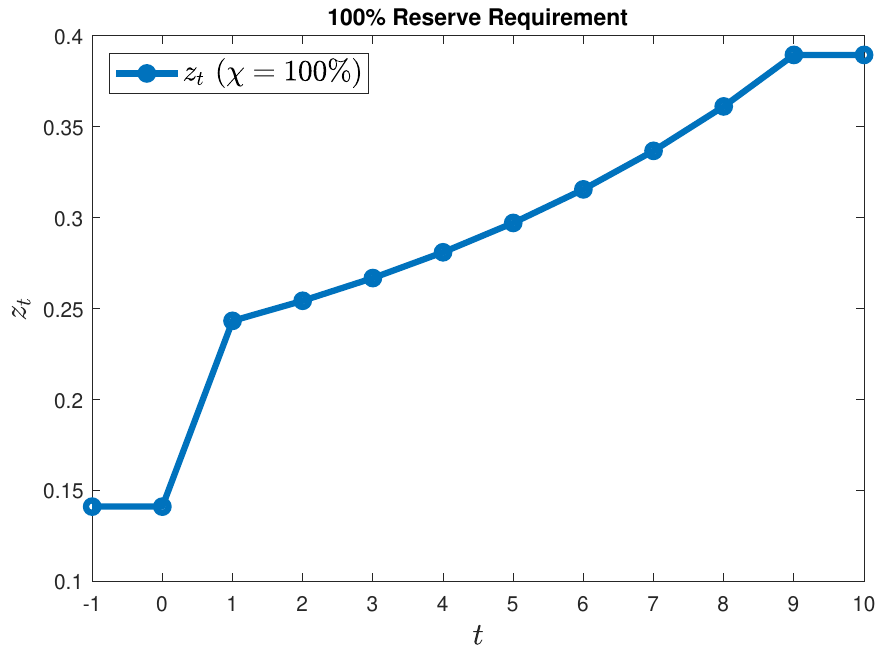}
\includegraphics[width=7cm,height=5cm]{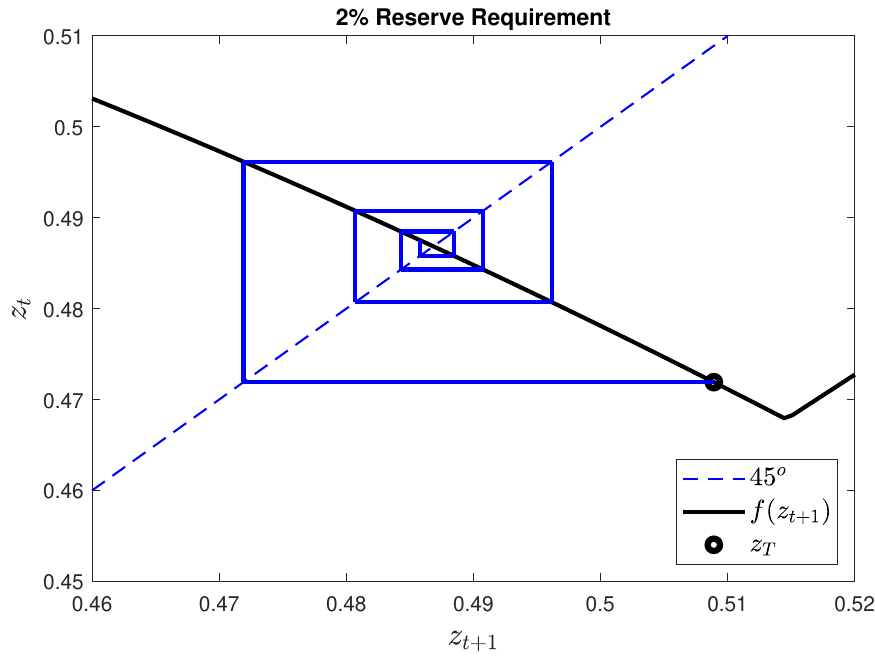}
\includegraphics[width=7cm,height=5cm]{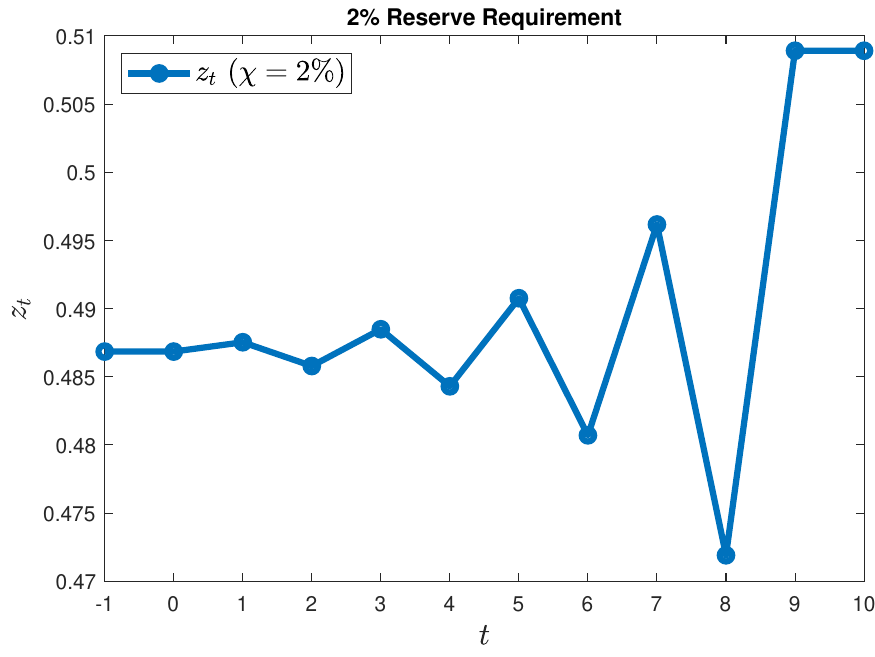}
\includegraphics[width=7cm,height=5cm]{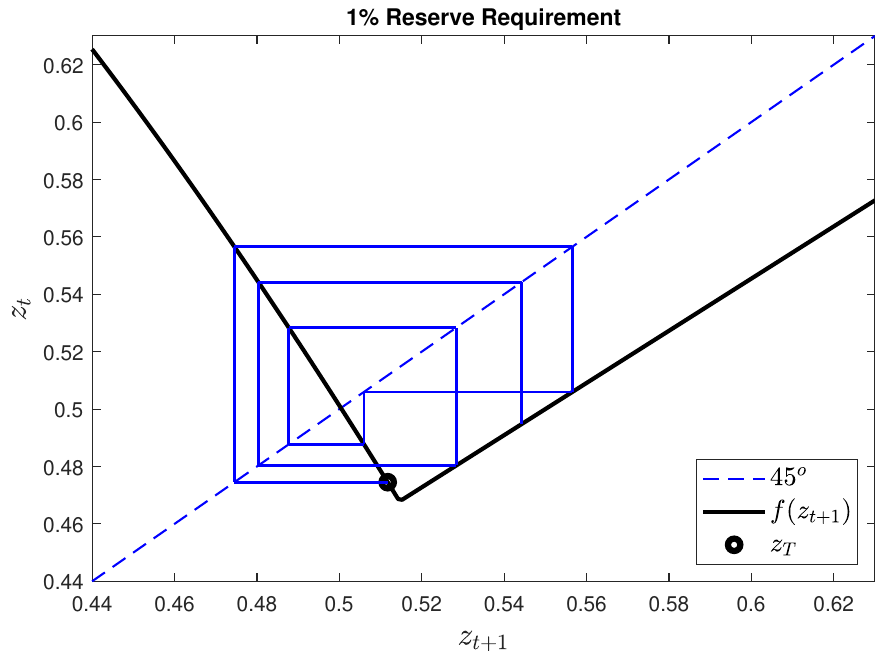}
\includegraphics[width=7cm,height=5cm]{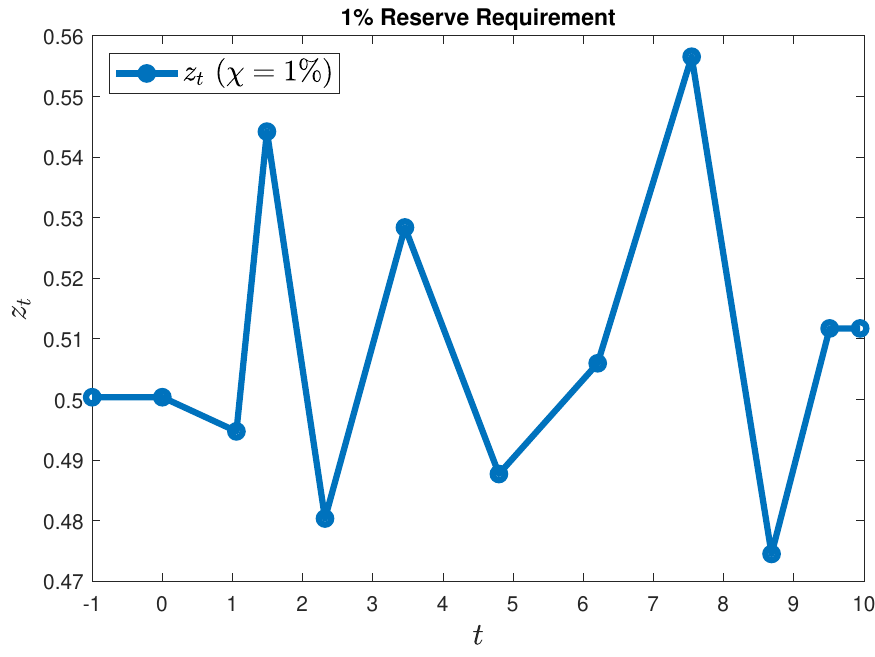}
\caption{Phase Dynamics and Transition Paths for Known Policy Change: Model 1}
\label{fig:trans}
\end{figure}

Consider a case with $T=9$. In Figure \ref{fig:trans}, we start with a stationary equilibrium with $i=0.1$. At time 0, it is announced that there will be a permanent change in $i$ to $0.02$ at time $T=9$. The stationary equilibrium with $i=0.02$ is a terminal condition. The top-right panel of Figure \ref{fig:trans} shows the transition path when $\chi=1$. It jumps right after the announcement and steadily converges to a new steady state. The top-left panel of Figures  \ref{fig:trans} displays its phase diagram. It does not have a backward bending feature, and $z_t$ increases monotonically until it reaches $z_T$.

The middle-right panel of Figure \ref{fig:trans} shows the transition path when $\chi=0.02$. Despite eventually reaching $z_T$, its transition dynamics are oscillatory. The middle-left panel of Figure \ref{fig:trans} presents the phase diagram. As discussed in Section \ref{sec:results}, when the reserve requirement is low, the function $f(\cdot)$ exhibits a backward-bending feature that induces oscillations on the path towards convergence to $z_T$. It is important to note that $\chi=0.02$ exceeds both $\chi_m$ and $\bar\chi_m$, implying that there is no endogenous two-period cycle in this case. However, even in the absence of an endogenous cycle, lower reserve requirements can still increase volatility in economic fluctuations when there is a news shock.

The bottom-right panel of Figure \ref{fig:trans} shows the transition path when $\chi=0.01$. This also exhibits an oscillatory transition. Even though it eventually goes up from $z_0$ to $z_T$,  it goes even higher than $z_T$ in the transition path.  The top-left panel of Figures  \ref{fig:trans} shows that its phase diagram. In this case, $f(\cdot)$ has a more backward bending feature than the case with $\chi=0.02$, and this feature induces large fluctuations during the path to converge to $z_T$.


\begin{figure}\centerfloat 
  \begin{subfigure}{0.5\textwidth}
    \includegraphics[width=\linewidth]{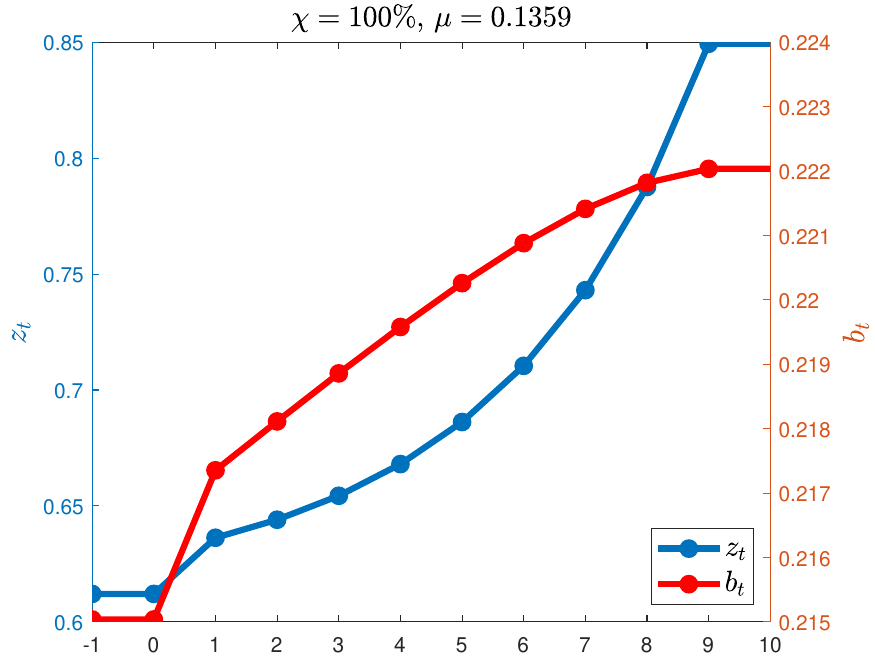}
    \caption{$\chi=1$} \label{fig:news_1}
  \end{subfigure}%
  \begin{subfigure}{0.5\textwidth}
    \includegraphics[width=\linewidth]{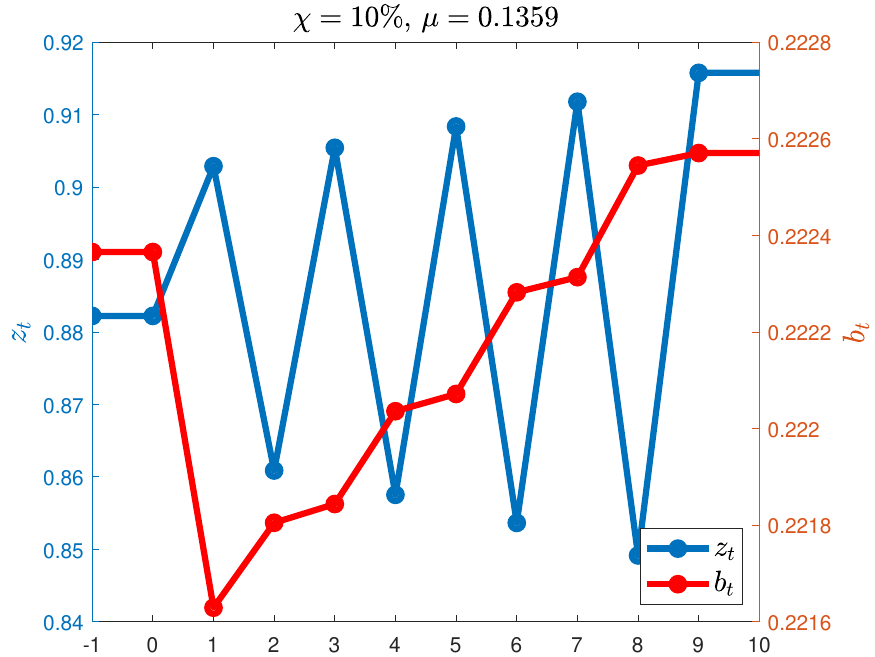}
    \caption{$\chi=0.1$} \label{fig:news_2}
  \end{subfigure}
\caption{Transition Paths for Known Policy Change: Model 2}
\label{fig:news}
\end{figure}

Now we introduce unsecured credit. Similar to the previous experiment, the permanent change from $i_0$ to $i_T$ implies a shift of equations (\ref{eq:mcdynamics1}) and (\ref{eq:mcdynamics2}), from $z_{t}=\Phi_0(z_{t+1},\bar{b}_{t+1};i_0,\chi)$ and $\bar{b}_{t}=\Gamma_0(z_{t+1},\bar{b}_{t+1};i_0,\chi)$ to $z_{t}=\Phi_T(z_{t+1},\bar{b}_{t+1};i_T,\chi)$ and $\bar{b}_{t}=\Gamma_T(z_{t+1},\bar{b}_{t+1};i_T,\chi)$, respectively. Then, starting in steady state with $(z_0,\bar{b}_0)$ and ending in steady state with $(z_T,\bar{b}_T)$, the transitional dynamics of the equilibrium with unsecured credit also can be solved by backward induction.
\begin{align*}
z_{T}=\Phi_T(z_{T},\bar{b}_{T}),\quad z_{T-1}=\Phi_0(z_{T},\bar{b}_{T}),\quad z_{T-2}&=\Phi_0(z_{T-1},\bar{b}_{T-1}), \quad...\quad z_{0}=\Phi_0(z_{1},\bar{b}_{1}) \\
\bar{b}_{T}=\Gamma_T(z_{T},\bar{b}_{T}),\quad \bar{b}_{T-1}=\Gamma_0(z_{T},\bar{b}_{T}),\quad \bar{b}_{T-2}&=\Gamma_0(z_{T-1},\bar{b}_{T-1}), \quad...\quad \bar{b}_{0}=\Gamma_0(z_{1},\bar{b}_{1})
\end{align*}

Again, let's consider a case with $T=9$. In Figure \ref{fig:news}, we begin with a stationary equilibrium with $i=0.1$. At time 0, an announcement is made that there will be a permanent change in $i$ to $0.02$ at time $T=9$. The stationary equilibrium with $i=0.02$ is a terminal condition for this dynamics system. Real money balances and credit converge monotonically to the new equilibrium when $\chi=1$ but they fluctuate considerably when reserve requirements are very low. Similar to Model 1, news about monetary policy induces complicated dynamics in $z$ and $b$ when reserve requirements are low. The right panel of Figure \ref{fig:news} also shows that there is an asymmetry when the reserve requirement is low. When $\chi=0.02$, news leading to an increase in $z_t$ tends to lower $b_t$ because higher $z_t$ raises equilibrium payoff, which in turn relaxes the endogenous credit limit.

As shown in Figure \ref{fig:trans} and \ref{fig:news}, transitions display various patterns depending on $\chi$, with lower reserve requirements more likely to induce cyclic and boom-bust responses. There is perfect foresight about the event and the transition is uniquely determined. However, it can display a wide range of patterns depending on reserve requirements. When reserve requirements are high, $z_t$ increases monotonically towards $z_T$. However, with low reserve requirements, the economy is more likely to experience cyclic and boom-bust responses following a monetary announcement.

\section{Empirical Evaluation: Inside Money Volatility}
\label{sec:empiric}

\begin{figure}[tp!]
\centering
\includegraphics[width=7cm,height=6cm]{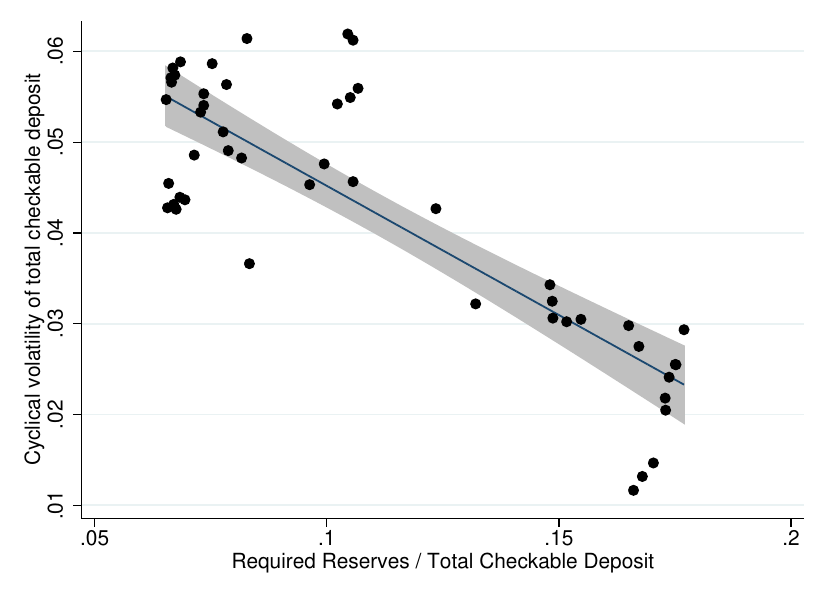}\label{fig:1}
\caption{Scatter Plot for Inside Money Volatility and Required Reserve Ratio}
\label{fig:scatter}
\end{figure}

In the previous sections, the theoretical results show that lowering the required reserve ratio can induce instability, as endogenous cycles are more likely to emerge with low reserve requirements. When endogenous cycles exist, lowering the reserve requirement increases the amplitude of the cycles. Also, the calibrated example shows that lowering reserve requirements increases the volatility of transition dynamics as the economy is more likely to induce cyclic and boom-bust responses from information shocks. Whether the economic fluctuation occurs from an exogenous shock or arises endogenously, the fractional reserve system increases the volatility of inside money real balances

To evaluate the model's prediction, I examine whether the required reserve ratio is associated with the cyclical volatility of the inside money real balance. Following \cite{jaimovich2009young} and \cite{carvalho2013great}, I measure the cyclical volatility in quarter $t$ as the standard deviation of a filtered log real total checkable deposit during a
41-quarter (10-year) window centered around quarter $t$. Total checkable deposits are from the H.6 Money Stock Measures published by the Federal Reserve Board and converted to real value using the Consumer Price Index (CPI). Seasonally adjusted series are used to smooth the seasonal fluctuation. I adopt the Hodrick-Prescott (HP) filter with a 1600 smoothing parameter as standard. To construct an annual series, quarterly observations are averaged for each year. The sample period is from 1960:I to 2018:IV so that there are annual series from 1965 to 2013 and quarterly series from 1965:I to 2013:IV.

\begin{table}[tp]
\centerfloat
\caption{Effect of Required Reserve Ratio}
\label{tab:reg1}
\footnotesize 
\begin{tabular}{lD{.}{.}{4}D{.}{.}{4}D{.}{.}{4}D{.}{.}{4}}
\toprule \hline
 & \multicolumn{2}{c}{Yearly}  & \multicolumn{2}{c}{Quarterly}  \\  \cline{2-3} \cline{4-5}
Dependent  & \multicolumn{1}{c}{OLS}  & \multicolumn{1}{c}{CCR}  & \multicolumn{1}{c}{OLS}  & \multicolumn{1}{c}{CCR}     \\ 
variable: $\sigma^{Roll}_t$  & \multicolumn{1}{c}{(1)}  & \multicolumn{1}{c}{(2)} & \multicolumn{1}{c}{(3)}  & \multicolumn{1}{c}{(4)}    \\ \hline
$\chi$        & -0.283^{***} & -0.245^{***} & -0.282^{***} & -0.452^{***}      \\ 
              & (0.027)      & (0.002)      & (0.016)      & (0.001)                \\
\texttt{ffr}  &              & -0.109^{***} &              & -0.050^{***}      \\
              &              & (0.002)      &              & (0.000)               \\
Constant      & 0.074^{***}  &  0.074^{***} & 0.074^{***}  &  0.085^{***}      \\
              & (0.003)      & (0.000)      & (0.002)      & (0.000)             \\
\hline
Obs. & \multicolumn{1}{c}{49} &\multicolumn{1}{c}{49}& \multicolumn{1}{c}{196} &\multicolumn{1}{c}{196}    \\
$adj R^2$ & \multicolumn{1}{c}{0.700} &\multicolumn{1}{c}{0.621} & \multicolumn{1}{c}{0.696} &\multicolumn{1}{c}{0.240}   \\ \hline
\multicolumn{5}{l}{\textbf{Johansen Tests for Cointegration}}   \\ 
$\lambda_{trace}(r=0)$ & \multicolumn{1}{c}{\textbf{9.81}}   & \multicolumn{1}{c}{35.69}  &\multicolumn{1}{c}{\textbf{10.93}}& \multicolumn{1}{c}{34.48}   \\
\text{ 5\% CV}         & \multicolumn{1}{c}{15.41} & \multicolumn{1}{c}{29.68}  & \multicolumn{1}{c}{15.41} & \multicolumn{1}{c}{29.68}  \\
$\lambda_{trace}(r=1)$ & \multicolumn{1}{c}{3.32}   &\multicolumn{1}{c}{\textbf{10.68} } & \multicolumn{1}{c}{1.94}  &\multicolumn{1}{c}{\textbf{12.09}}    \\
\text{ 5\% CV}         & \multicolumn{1}{c}{3.76}  & \multicolumn{1}{c}{15.41}   & \multicolumn{1}{c}{3.76} & \multicolumn{1}{c}{15.41}   \\ \hline\bottomrule   
\end{tabular}
\begin{center}
\begin{minipage}{0.9\textwidth} 
{\footnotesize Note: For (1), (3), (5) and (7), OLS estimates are reported, and Newey-West standard errors with lag 1 are reported in parentheses. For (2), (4), (6), and (8), first-stage long-run variance estimations for CCR are based on the quadratic spectral kernel and Bayesian information criterion. The bandwidth selection is based on Newey-West fixed lag, $4\times(T/100)^{2/9}$; $\chi$ denotes the required reserve ratio, \texttt{ffr} denotes federal funds rates and $\sigma^{Roll}_t$ denotes the cyclical volatility of real inside money balances. ***, **, and * denotes significance at the 1, 5, and 10 percent levels, respectively. \par}
\end{minipage}    
\end{center}
\end{table}

To compute the required reserve ratios, I divide the required reserves by the deposit component of M1 instead of using the official legal reserve requirement. The reason for this approach is the following. The legal reserve requirement for demand deposits was 10\% from April 2, 1992, to March 25, 2020. However, the Federal Reserve imposed different reserve requirements depending on the size of a bank's liabilities. For example, from December 29, 2011, to December 26, 2012, the Fed had a reserve requirement exemption for liabilities up to \$11.5 million. For liabilities between \$11.5 million and \$71.0 million, the Fed imposed a 3\% reserve requirement. These criteria have changed over time. On December 27, 2012, the Fed increased the exemption threshold to \$12.4 million and raised the low reserve tranche from \$71.0 million to \$79.5 million. During 1992:I-2019:IV, there were 27 changes in these thresholds. Dividing the required reserves by the deposit component of M1 allows us to track these changes as well. 

\begin{table}[tp]
\centering
\small 
\caption{Unit Root Tests}
\label{tab:unit_root}
\begin{tabular}{llD{.}{.}{6}D{.}{.}{6}D{.}{.}{6}}
\toprule \hline
& &  \multicolumn{2}{c}{Phillips-Perron test} & \multicolumn{1}{c}{ADF test}  \\ \hline
& & \multicolumn{1}{c}{$Z(\rho)$} & \multicolumn{1}{c}{$Z(t)$} &
\multicolumn{1}{c}{$Z(t)$} \\ \hline
\textbf{Yearly} & \texttt{ffr}      & -6.766 & -1.704 & -2.362  \\ 
& $\chi$            & -1.492 & -1.173 & -1.341  \\ 
& $\sigma^{Roll}_t$ & -4.708 & -2.191 & -2.090 \\ 
 \cline{2-5}
& $\Delta\texttt{ffr}$ & -28.373^{***}  & -5.061^{***}  &-6.357^{***}  \\ 
& $\Delta\chi$          & -31.818^{***}  & -4.802^{***}  &-3.693^{***}   \\
& $\Delta\sigma^{Roll}_t$ & -24.905^{***}  & -3.416^{**}  &-2.942^{**}    \\ \hline 
\textbf{Quarterly} & \texttt{ffr}         & -8.611  & -1.956 & -2.183  \\ 
& $\chi$               & -1.335  & -1.145 & -1.199  \\
& $\sigma^{Roll}_t$    & -4.320  & -2.062 & -1.554 \\   \cline{2-5}
& $\Delta\texttt{ffr}$ & -139.701^{***} & -10.792^{***} & -10.288^{***}  \\ 
& $\Delta\chi$         & -163.796^{***} & -12.272^{***} & -9.909^{***}   \\
& $\Delta\sigma^{Roll}_t$ & -23.132^{***}  & -2.604^{*}    & -3.576^{***} \\  
\hline \bottomrule
\end{tabular}
\begin{center}
\begin{minipage}{0.8\textwidth} 
{\footnotesize Note: \texttt{ffr} denotes federal funds rates, $\chi$ denotes required reserve ratio, and $\sigma^{Roll}_t$ denotes cyclical volatility of real inside money balances. All series are demeaned before implementing the unit root test following to \cite{elliott2006minimizing} and \cite{harvey2009unit}, because the magnitude of the initial value can be problematic. Let ***, **, and * denotes significance at the 1, 5, and 10 percent levels, respectively. \par}
\end{minipage}    
\end{center}
\end{table}

Figure \ref{fig:scatter} presents a scatter plot of the cyclical volatility of the real inside money balance and the required reserve ratio. Columns (1) and (3) of Table \ref{tab:reg1} reports its regression estimates with Newey-West standard errors. The plot and estimates show a negative relationship between the cyclical volatility of the real inside money balance and the required reserve ratio with statistically significant regression coefficients. However, this result can be driven by a spurious regression. Table \ref{tab:unit_root} provides unit root test results for the federal funds rate, the required reserve ratio, and the cyclical volatility of inside money. Both augmented Dickey-Fuller tests and Phillips-Perron tests fail to reject the null hypotheses of unit roots for these series, whereas they reject the null hypotheses of unit roots at their first differences. In addition to that, the Johansen cointegration test in Columns (1) and (3), suggests that there is no cointegration relationship between two variables. So it is hard to rule out that the results of Columns (1) and (3) are driven by a spurious regression.

To overcome this issue, I adopt the cointegrating regression with an additional variable, the federal funds rate. Columns (2) and (4) of Table \ref{tab:reg1} provide Johansen cointegration test results for the federal funds rate, the required reserves, and the cyclical volatility of inside money. The trace test suggests a cointegration relationship among these three variables, which is consistent with the theoretical result: The instability depends on the reserve requirement and the interest rate. With the cointegration relationship, we may not have to worry about a spurious relationship. Columns (2) and (4) of Table \ref{tab:reg1} report the estimates for the cointegrating relationship. Because of the potential bias from long-run variance, I estimate a canonical cointegrating regression (CCR). The estimates are statistically significant with a sizable effect. The cointegration analysis confirms that a lower reserve requirement is associated with higher volatility of inside money real balances, consistent with the model's prediction.

\section{Conclusion}
\label{sec:conclusion}
The goal of this paper is to examine the (in)stability of fractional reserve banking. To that end, this paper builds a simple monetary model of fractional reserve banking that can capture the conditions for (in)stability under different specifications. Lowering the reserve requirement increases the consumption at the steady state. However, it can induce instability. The baseline model and its extension establish the conditions for endogenous cycles and chaotic dynamics. The model also features stochastic cycles and self-fulfilling boom and burst under explicit conditions. The model shows that fractional reserve banking can endanger stability in the sense that equilibrium is more prone to exhibit cyclic, chaotic, and stochastic dynamics under lower reserve requirements. This is due to the amplified liquidity premium. This result holds in the extended model with unsecured credit. 

The calibrated exercise suggests that this channel could be another source of economic fluctuations. This paper also provides some empirical evidence that is consistent with the prediction of the model. I test the association between the required reserves ratio and the real money volatility using cointegrating regression. I find a significant negative relationship between the two variables.   Both theoretical and empirical evidence find a link between the reserve requirement policy and (in)stability.

\newpage
\bibliography{bibi}
\bibliographystyle{aer} 

\newpage
\textbf{\huge{Online Appendix}}

\begin{appendices}

\section{Proofs}

\label{sec:proof}

\begin{proof}[Proof of Proposition \ref{prop:welfare}]
Recall (\ref{eq:mss})
$$\chi i=(1-\sigma+\sigma\chi)\alpha L(z_s).$$
Since $L'(\cdot)<0$, we have the following:
$$\frac{\partial z_s}{\partial i }=\frac{\chi}{(1-\sigma+\sigma\chi)\alpha L'(z_s)}<0, \quad \text{ and }\quad\frac{\partial z_s}{\partial \chi}=\frac{i-\sigma\alpha L(z_s)}{(1-\sigma+\sigma\chi)\alpha L'(z_s)}<0. $$
Since $z=v(q)$ and $v'(q)>0$, it is straightforward to show that lowering $i$ or lowering $\chi$ increases $q$.
\end{proof}

\begin{proof}[Proof of Proposition \ref{prop:2cycle}]
Let there exists a two-period cycle satisfying  $z_1<z_s<p^*\leq z_2$. Since $z_2\geq p^*$, we have $z_2=(1+i)z_1$. Using (\ref{eq:mdynamics}) with $z_1<p^*$ gives
\begin{equation}
\chi=\frac{(1-\sigma)\alpha L(z_1) }{(1+i)^2-1-\sigma\alpha L(z_1)} 
\end{equation}
This two-period cycle should satisfy $z_1<z_s<p^*$ and $z_2=(1+i)z_1\geq p^*$. The first one can be easily shown using  
$$0=L(p^*)<L(z_s)=\frac{i}{\alpha(1-\sigma+\sigma\chi)}\chi < \frac{(1+i)^2-1}{\alpha(1-\sigma+\sigma\chi)}\chi=L(z_1)$$
since we have $L'(\cdot)<0$. Because $dz_1/d\chi<0$, the latter one, $z_1\geq p^*/(1+i)$, is held when
$$0<\chi\leq\frac{(1-\sigma)\alpha L\left(\frac{p^*}{1+i}\right) }{(1+i)^2-1-\sigma\alpha L\left(\frac{p^*}{1+i}\right)}.$$
This equilibrium solves 
$$\frac{(1+i)^2-1}{\alpha(1-\sigma+\sigma\chi)}\chi=L(z_1), \text{ and } z_2=(1+i)z_1$$
We can check if lowering the reserve requirement also increases the volatility. Consider the difference between peak and trough $z_2 -z_1=iz_1$. Since 
$$\frac{\partial z_1}{\partial \chi} =\frac{\alpha(1-\sigma)}{\chi\{(1+i)^2-1\}}\frac{\{L(z_1)\}^2}{L'(z_1) }<0,$$ 
reducing the reserve requirement increases the difference between peak and trough. 
\end{proof}

\begin{proof}[Proof of the Existence of a Two-period Monetary Cycle where $f'(z)<-1$]
Let $f^2(z)=f\circ f(z)$.  With given the unique steady state, $f(z)>z$ for $z<z_s$ and $f(z)<z$ for $z>z_s$.  Because $f(z)$ is linear increasing function for $z>p^*$, there exist a $\tilde{z}>p^*$ s.t $f(\tilde{z})>p^*$. Since $\tilde{z}>p^*$ and $f(\tilde{z})<\tilde{z}$, $\tilde{z}$ satisfies $f^2(\tilde{z})<f(\tilde{z})<\tilde{z}$. We can write slope of $f^2(z)$ as follows. 
$$\frac{\partial f^2(z)}{\partial z}=f'[f(z)]f'(z)=f'(z)f'(z)=[f'(z)]^2$$
which implies $\partial f^2(z)/\partial z >1$ when  $f'(z)<-1$. And it is easy to show $\partial f^2(0)/\partial z>0$. With  given $i>0$ and $\chi>0$, there exist a $(z_1,z_2)$, satisfying $0<z_1<z_s<z_2$ which are fix points for $f^2(z)$.
\end{proof}

\begin{proof}[Proof of Proposition \ref{prop:speical}]
When DM trade is based on take-it-or-leave-it offer from buyer to seller with $c(q)=q$ and  $-qu''(q)/u'=\eta$, $f'(q)$ can be written as
$$f'(q)=\frac{1}{1+i}\left\{\frac{1-\sigma+\sigma\chi}{\chi}\alpha\left[
u''(q)q+u'(q)-1\right]+1\right\}<-1$$
Using $u''(q)q=-\eta u'(q)$ gives

$$\frac{1-\sigma+\sigma\chi}{\chi}\alpha\left[
u'(q)(1-\eta)-1\right]+1<-(1+i)$$
where $u'(q)=1+\frac{i\chi}{\alpha(1-\sigma+\sigma\chi)}$. Substituting $u'(q)$ yields 
$$
\left\{ 1+\frac{i\chi}{\alpha(1-\sigma+\sigma\chi)}  \right\}(1-\eta)-1<-\frac{\chi(2+i)}{\alpha(1-\sigma+\sigma\chi)}. $$
Then rearranging terms gives 
$$ 0<\chi<\frac{\alpha\eta(1-\sigma)}{\eta(1-\alpha\sigma)+(2-\eta)(1+i)}. $$
\end{proof}

\begin{proof}[Proof of Proposition \ref{prop:3cycle}]
I divide three period cycles into two cases. \\
Case 1: Let there exists a three-period cycle satisfying  $z_1<z_s<p^*\leq z_2<z_3$. Since $z_2,z_3\geq p^*$, we have $z_2=(1+i)z_1$, $z_3=(1+i)z_2=(1+i)^2 z_1$. Using (\ref{eq:mdynamics}) with $z_1<p^*$ gives
\begin{equation}
\chi=\frac{(1-\sigma)\alpha L(z_1) }{(1+i)^3-1-\sigma\alpha L(z_1)} 
\end{equation}
This three-period cycle should satisfy $z_1<z_s<p^*$ and $z_2=(1+i)z_1\geq p^*$. First one can be easily shown using  
$$0=L(p^*)<L(z_s)=\frac{i}{\alpha(1-\sigma+\sigma\chi)}\chi < \frac{(1+i)^3-1}{\alpha(1-\sigma+\sigma\chi)}\chi=L(z_1)$$
since we have $L'(\cdot)<0$. Because $dz_1/d\chi<0$, the latter one, $z_1\geq p^*/(1+i)$, is held when
$$0<\chi \leq \frac{(1-\sigma)\alpha L\left(\frac{p^*}{1+i}\right) }{(1+i)^3-1-\sigma\alpha L\left(\frac{p^*}{1+i}\right)}.$$
Case 2: Let there exists a three-period cycle satisfying  $z_1<z_2<p^*\leq z_3$. Since $z_3>p^*$, we have $z_3=z_2(1+i)$ and $(z_2, z_1)$ solves (\ref{eq:axc11})-(\ref{eq:axc12}). 
\begin{align}
z_1&=f(z_2)=\left[\frac{1-\sigma+\sigma\chi}{\chi}\alpha L(z_2)+1 \right]\frac{z_2}{1+i} \label{eq:axc11} \\
z_2&= \tilde{f}(z_1)\equiv\left[\frac{1-\sigma+\sigma\chi}{\chi}\alpha L(z_1)+1  \right]\frac{z_1}{(1+i)^2}\label{eq:axc12}.
\end{align}
\begin{figure}[h!]
\centering
\includegraphics[width=7.5cm,height=6cm]{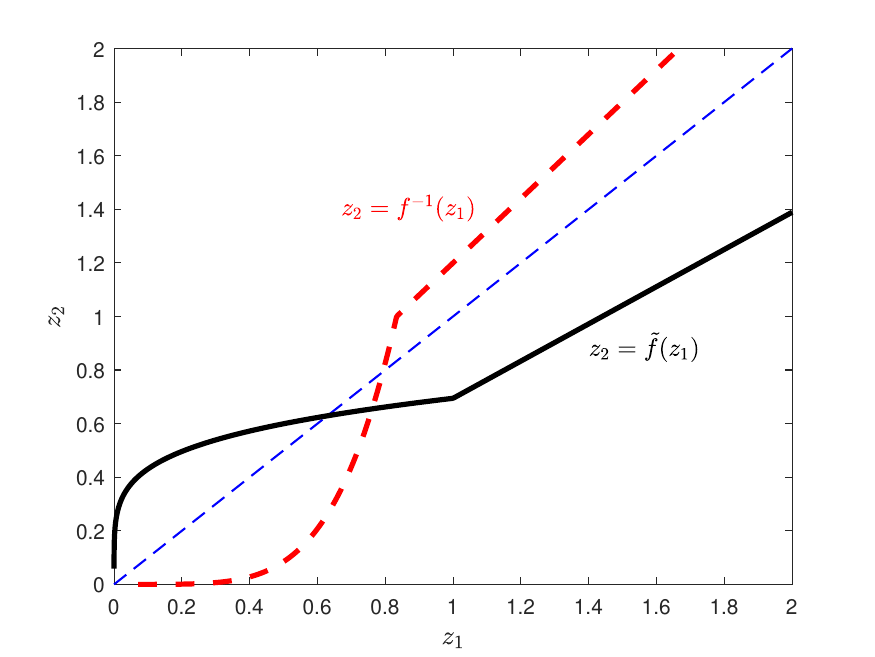}
\includegraphics[width=7.5cm,height=6cm]{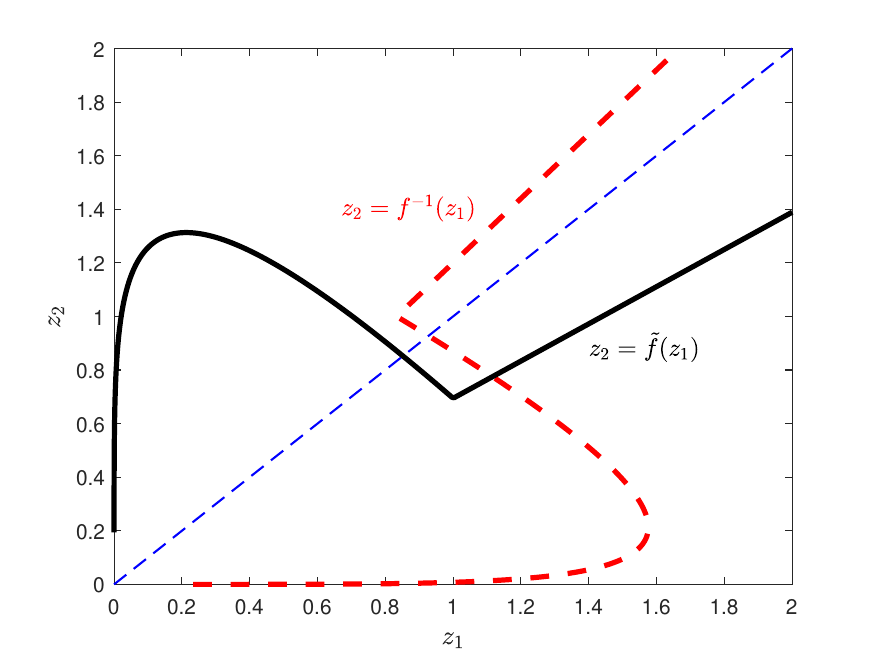}
\caption{Intersection of $\tilde{f}(z)$ and $f(z)$}
\end{figure} 
These functions satisfies $f(x)>x$ for $x<z_s$, $f(x)<x$ for $x>z_s$, $\tilde{f}(x)>x$ for $x<\tilde{z}$ and $\tilde{f}(x)<x$ for $x>\tilde{z}$ where 
$\tilde{z}$ solves $\tilde{z}=\tilde{f}(\tilde{z})$. One can easily show $\tilde{z}<z_s$. Therefore any intersection between $z_1=f(z_2)$ and $z_2=\tilde{f}(z_1)$ satisfies $z_1>z_2$ which contradicts to our initial conjecture $z_1<z_2$. This implies there is no three-period cycle satisfying  $z_1<z_2<p^*\leq z_3$. 
Therefore we can conclude that a three-period cycle exists when
$$0<\chi\leq\frac{(1-\sigma)\alpha L\left(\frac{p^*}{1+i}\right)}{(1+i)^3-1-\sigma\alpha L\left(\frac{p^*}{1+i}\right)}.$$
This equilibrium solves 
$$\frac{(1+i)^3-1}{\alpha(1-\sigma+\sigma\chi)}\chi=L(z_1),\quad z_2=(1+i)z_1, \quad \text{ and } z_3=(1+i)z_2.$$
We can check if lowering the reserve requirement also increases the volatility. Consider the difference between peak and trough $z_3 -z_1=(i^2+2i)z_1$. Since 
$$\frac{\partial z_1}{\partial \chi} =\frac{\alpha(1-\sigma)}{\chi\{(1+i)^3-1\}}\frac{\{L(z_1)\}^2}{L'(z_1) }<0,$$ 
reducing the reserve requirement increases the difference between peak and trough.

The existence of a three-cycle
implies the existence of cycles of all orders and chaotic dynamics by the Sarkovskii theorem \citep{sharkovskii1964cycles} and the Li-Yorke theorem \citep{li1975period}.
\end{proof}

\begin{proof}[Proof of Corollary \ref{coro:least}:]
Proposition \ref{prop:3cycle} shows that at least one periodic point satisfies $z_t<z_s<p^*$ in 3- period cycles. Two period cycles satisfies $z_1<z_s<z_2$ also implies at least one periodic point satisfies $z_t<z_s<p^*$ in 2-period cycles since $z_1<z_s<p^*$. This result holds for any $n$-periodic cycles. Let $z_1<z_2<...z_n$ be the periodic points of a $n$-cycle. Suppose $z_j>z_s$ for all $j=1,2,..n$. By the definition of a $n$-period cycle, $z_1=f(z_n)<z_n$ since  $f(z)<z$ for $z>z_s$.
$$z_n=f(z_{n-1})<z_{n-1}=f(z_{n-2})<z_{n-2}...<z_1.$$ which shows the contradiction implying at least one periodic point satisfies $z_t<z_s<p^*$.  \end{proof}

\begin{proof}[Proof of Proposition \ref{prop:sunspot}]
By definition, if there exists $(\zeta_1,\zeta_2)$ satisfying
\begin{align} 
z^1&=\zeta_1 f(z^1)+(1-\zeta_1)f(z^2) \label{eq:sse_1}   \\
z^2&=(1-\zeta_2)f(z^1)+\zeta_2 f(z^2)  \label{eq:sse_2}  
\end{align}
with $\zeta_1,\zeta_2<1$, then there exists a proper sunspot equilibrium.  Because $z^1$ and $z^2$ are weighted averages of $f(z^1)$ and $f(z^2)$, where $f(z^1)>z^1$ and $f(z^2)<z^2$, by the uniqueness of the positive steady state, necessary and sufficient conditions for (\ref{eq:sse_1})  and (\ref{eq:sse_2}) are
$$f(z^2)<z_1<f(z^1) \text{ and } f(z^2)<z_2<f(z^1).$$
Since $z^1<z^2$, above conditions are reduce to 
\begin{align} \label{eq:suff}
z^2<f(z^1) \text{ and } z^1>f(z^2).
\end{align}
When $\chi<\chi_m$, there exists $(z^1,z^2)$ that satisfies (\ref{eq:suff}). 
Rewrite (\ref{eq:sse_1})  and (\ref{eq:sse_2}) as 
\begin{align} \label{eq:sunspot_proof}
\zeta_1+\zeta_2=\frac{z^1-f(z^2)-z^2+f(z^1)}{f(z^1)-f(z^2)}=\frac{z^1-z^2}{f(z^1)-f(z^2)}+1<1
\end{align}
since $z^2<z^1$ and $f(z^1)>f(z^2)$. Therefore, when $\chi<\chi_m$, a stationary sunspot equilibrium exists. 

Now consider the case with $f'(z_s)<-1$. Since $f'(z_s)<0$, there is an interval $[z_s-\varepsilon_1,z_s+\varepsilon_2 ]$,  which satisfy $\varepsilon_1,  \varepsilon_2>0$ and $f(z^1)>f(z^2)$ for $z^1\in[z_s-\varepsilon_1,z_s)$ and $z^2\in(z_s,z_s+\varepsilon_2]$.
$$\frac{z^2-z_s}{z_s-z^1}< -f'(z_s)<\frac{z_s-z^1}{z^2-z_s}$$
Since  $f'(z_s)<-1$, the above condition can be reduce to $-f'(z_s)<\frac{z_s-z^1}{z^2-z_s}=\frac{\varepsilon_1}{\varepsilon_2}$. There exist multiple solutions, $(\varepsilon_1,  \varepsilon_2)$, satisfying $-f'(z_s)\varepsilon_2<\varepsilon_1$ given $-f'(z_s)>1$ and $\varepsilon_1,\varepsilon_2>0$. These solutions satisfy (\ref{eq:sunspot_proof}). Therefore, if $f'(z_s)<-1$, there exists a stationary sunspot cycle.
\end{proof}

\begin{proof}[Proof of Proposition \ref{prop:burst}]
Consider $z_{t}=f(z_{t+1})$. If $z_s>\bar z$ where  $\bar z$ solves $f'(\bar z)=0$. In this case, there exist multiple equilibria. If  $q^*\leq f(\bar z)$,  then there exist equilibria $\{z_t\}_{t=0}^\infty$  with $z_T\equiv\max \{z_t\}^{\infty}_{t=0}>q^*$ (bubble) which crashes to 0 (burst) as $t\rightarrow\infty$, where $T\geq 1$ and $z_T>z_0$. Then there exist equilibria with bubble-burst as a self-fulfilling crisis. Conditions for this case are shown as below. 
Similar to Corollary \ref{prop:speical}, consider take-it-leave-it offer with $-q u''/u'=\eta$ and $c(q)=q$. Then we have following difference equation: 
\begin{align}
z_{t}&=f(z_{t+1})
\begin{dcases} 
\frac{z_{t+1}}{1+i}\left\{\frac{1-\sigma+\sigma\chi}{\chi}\alpha\left[u'(z_{t+1})-1\right]+1\right\} &\text{ if } z_{t+1}< q^* \\
\frac{z_{t+1}}{1+i}             &\text{ if } z_{t+1}\geq q^*
\end{dcases}
\end{align}

Step 1: [Multiplicity i.e., $z_s >\bar z$ where $\bar{z}$ solves $f'(\bar{z})=0$] Consider the following condition. 
$$f'(\bar{z})=\frac{1}{1+i}\left\{\frac{\alpha(1-\sigma+\sigma\chi)}{\chi}[u'(\bar{z})(1-\eta)-1]+1\right\}=0$$
Since $z_s>\bar{z} \rightarrow u'(z_s)<u'(\bar{z})$, we have 
$$u'(z_s)=1+\frac{i\chi}{\alpha(1-\sigma+\sigma\chi)}<\frac{1}{1-\eta}\left\{1- 
\frac{\chi}{\alpha(1-\sigma+\sigma\chi)}\right\}
=u'(\bar{z}).$$
This can be reduced as 
$$\chi<\frac{\alpha\eta(1-\sigma)}{1+i-\eta(i+\alpha\sigma)}$$
Step 2: [Show $q^*\leq f(\bar{z})$] It is straightforward to show that $q^*<f(\bar{z})$ holds when
$$\chi<\frac{(1-\sigma)\alpha\eta(1+i)}{(1-\eta)^2 q^*+(1+i)[(1-\eta)(3+i-\eta)-\alpha\sigma\eta]}$$
Therefore, when 
$$0<\chi<\min\left\{\frac{(1-\sigma)\alpha\eta(1+i)}{(1-\eta)^2 q^*+(1+i)[(1-\eta)(3+i-\eta)-\alpha\sigma\eta]}, \frac{\alpha\eta(1-\sigma)}{1+i-\eta(i+\alpha\sigma)}\right\}$$
there exist $\{z_t\}_{t=0}^{\infty}$ satisfying $z_T\equiv\max \{z_t\}^{\infty}_{t=0}>q^*$ and $\lim_{t\rightarrow\infty}z_t=0$, where $T\geq 1$ and $z_T>z_0>q^*/(1+i)$.
\end{proof}

\begin{proof}[Proof of Proposition \ref{prop:mc3cycle}]
A two period cycle result is presented and three-period case will follow. Let there exists a two-period cycle satisfying  $w_1<q^*<w_2$ where $w_j=z_j+\bar{b}_j$. Since $w_2>q^*$, we have $z_2=(1+i)z_1$ and $\bar{b}_2=(1+r)\bar{b}_1$ where $q_1$, $\bar{b}_1$, and $z_1$ solve
\begin{align*}
u'(q_1)&=1+\chi\frac{(1+i)^2-1}{\alpha(1-\sigma+\sigma\chi)}\\
\bar{b}_1&=
[(1+\rho)^2-1]^{-1}\left\{
\frac{i\mu\sigma\chi}{1-\sigma+\sigma\chi}\left[1-\frac{(1+i)^2}{\beta}\right]z_1+\mu\alpha\sigma [u(q_1)-q_1]
\right\}
\end{align*}
and $z_1=q_1-\bar{b}_1$.  
This two-period cycle should satisfy $q_1<q^*$ and $w_2=(1+i)z_1+(1+r)\bar{b}_1>q^*$. For given $i>0$ and $\chi>0$, first one can be easily shown using  
$$1=u'(q^*)<u'(q_s)=1+\frac{i}{\alpha(1-\sigma+\sigma\chi)}\chi < 1+\frac{(1+i)^2-1}{\alpha(1-\sigma+\sigma\chi)}\chi=u'(q_1)$$
since we have $u''(\cdot)<0$. 
Now we also can check the latter using the below conditions
\begin{align*}
(1+\rho)q_1>&(1+i)z_1+(1+\rho)\bar{b}_1=w_2>q^*>q_1=z_1+\bar{b}_1   \quad \text{ if }  \rho>i\\
(1+i)q_1>&(1+i)z_1+(1+\rho)\bar{b}_1=w_2>q^*>q_1=z_1+\bar{b}_1 \quad \text{ if } i>\rho.
\end{align*}
The sufficient conditions to have $w_2>q^*$ is $q_1>q^*/(1+\rho)$ for $\rho>i$ and  $q_1>q^*/(1+i)$ for $i>\rho$. Since we have $dq_1/d\chi<0$, there exist a three period cycle $q_1=w_1<q_s<q^*<w_2<w_3$ when  
$$0<\chi<\frac{(1-\sigma)\alpha [u'\left(\frac{q^*}{1+\iota}\right) -1]}{(1+i)^2-1-\sigma\alpha [u'\left(\frac{q^*}{1+\iota}\right) -1]}$$
where $\iota=\max\{i,\rho\}$. 
Now, let there exists a three-period cycle satisfying  $q_1=w_1<q_s<q^*<w_2<w_3$ where $w_j=z_j+\bar{b}_j$. Since $w_3$, $w_2>q^*$, we have $z_2=(1+i)z_1$, $z_3=(1+i)^2z_1$, $\bar{b}_2=(1+\rho)\bar{b}_1$  and $\bar{b}_3=(1+\rho)^2\bar{b}_1$ where $q_1$, $\bar{b}_1$, and $z_1$ solve 
\begin{align*}
u'(q_1)&=1+\chi\frac{(1+i)^3-1}{\alpha(1-\sigma+\sigma\chi)}    \\
\bar{b}_1&=
[(1+\rho)^3-1]^{-1}\left\{
\frac{i\mu\sigma\chi}{1-\sigma+\sigma\chi}\left[1-\frac{(1+i)^2}{\beta}\right]z_1+\mu\alpha\sigma [u(q_1)-q_1]
\right\}
\end{align*}
and $z_1=q_1-\bar{b}_1$.  
This three-period cycle should satisfy $q_1<q_s<q^*$ and $w_2=(1+i)z_1+(1+\rho)\bar{b}_1>q^*$. For given $i>0$ and $\chi>0$, first one can be easily shown using  
$$1=u'(q^*)<u'(q_s)=1+\frac{i}{\alpha(1-\sigma+\sigma\chi)}\chi < 1+\frac{(1+i)^3-1}{\alpha(1-\sigma+\sigma\chi)}\chi=u'(q_1)$$
since we have $u''(\cdot)<0$. 
Now we also can check the latter using below conditions
\begin{align*}
(1+\rho)q_1>&(1+i)z_1+(1+\rho)\bar{b}_1=w_2>q^*>q_1=z_1+\bar{b}_1   \quad \text{ if }  \rho>i\\
(1+i)q_1>&(1+i)z_1+(1+\rho)\bar{b}_1=w_2>q^*>q_1=z_1+\bar{b}_1 \quad \text{ if } i>\rho.
\end{align*}
The sufficient conditions to have $w_2>q^*$ is $q_1>q^*/(1+\rho)$ for $\rho>i$ and  $q_1>q^*/(1+\rho)$ for $i>\rho$. Since we have $dq_1/d\chi<0$, there exist a three period cycle $q_1=w_1<q_s<q^*<w_2<w_3$ when  
$$0<\chi<\frac{(1-\sigma)\alpha [u'\left(\frac{q^*}{1+\iota}\right) -1]}{(1+i)^3-1-\sigma\alpha [u'\left(\frac{q^*}{1+\iota}\right) -1]}$$
where $\iota=\max\{i,\rho\}$. Again, the existence of a three-cycle
implies the existence of cycles of all orders and chaotic dynamics by the Sarkovskii theorem and the Li-Yorke theorem.
\end{proof}

\end{appendices}

\end{document}